\documentclass[11pt]{article}

\usepackage{amsfonts}
\usepackage{amsmath}
\usepackage{amssymb}
\usepackage{amsthm}
\usepackage[mathscr]{eucal}
\usepackage{bbold}
\usepackage{braket}
\usepackage{color}
\usepackage{dsfont}
\usepackage{framed}
\usepackage{jheppub}
\usepackage{mathtools}
\usepackage{physics}
\usepackage[normalem]{ulem}
\usepackage{tensor}
\usepackage{thmtools}
\usepackage{thm-restate}
\usepackage{ulem}
\usepackage{tikz}
\usetikzlibrary{math} 

\definecolor{rcolor}{RGB}{200,0,0}

\definecolor{gcolor}{RGB}{0,100,0}

\definecolor{bcolor}{RGB}{10,10,255}

\definecolor{ocolor}{RGB}{255,165,0}

\definecolor{pcolor}{RGB}{255,10,255}

\newcommand{\ignore}[1]{}

\newtheorem*{ex}{Exercise}

\newcommand{\Eqref}[1]{Eq.~\eqref{#1}}
\newcommand{\secref}[1]{Sec.~\ref{#1}}
\newcommand{\figref}[1]{Fig.~\ref{#1}}

\makeatletter\def\@fpheader{~}\makeatother
\newcommand{\rnum}[1]
    {\MakeUppercase{\romannumeral #1}}
\makeatletter
\newcommand*{\defeq}{\mathrel{\rlap{%
                     \raisebox{0.3ex}{$\m@th\cdot$}}%
                     \raisebox{-0.3ex}{$\m@th\cdot$}}%
                     =} 
\makeatother

\def\be{\begin{eqnarray}}
\def\ee{\end{eqnarray}}

\newcommand{\bea}{\begin{eqnarray}}
\newcommand{\eea}{\end{eqnarray}}
\def\ben{\begin{equation}}
\def\een{\end{equation}}

\let\z=\zeta   
\let\l=\lambda     \let\r=v

\let\f=\frac

\let\pa=\partial
\def\be{\begin{equation}}
\def\ee{\end{equation}}
\def\ba{\begin{eqnarray}}
\def\ea{\end{eqnarray}}

\newcommand{\zb}{\bar{z}}
\def\bal#1\eal{\begin{align}#1\end{align}}
\def\bs#1\es{\begin{split}#1\end{split}}

\allowdisplaybreaks  

\numberwithin{equation}{section}

\def\be{\begin{equation}}
\def\ee{\end{equation}}
\def\ba{\begin{eqnarray}}
\def\ea{\end{eqnarray}}
\def\bal#1\eal{\begin{align}#1\end{align}}

\def\r{\rightarrow}

\def\f {\frac}

\def\l{\left}
\def\r{\right}

\def\z{\bar{z}}

\title{The Spacetime Geometry of Fixed-Area States in Gravitational Systems}

\author{Xi Dong,} \emailAdd{xidong@ucsb.edu} 
\author{Donald Marolf,} \emailAdd{marolf@ucsb.edu} 
\author{Pratik Rath,} \emailAdd{rath@ucsb.edu} 
\author{Amirhossein Tajdini and} \emailAdd{ahtajdini@ucsb.edu} 
\author{Zhencheng Wang} \emailAdd{zhencheng@ucsb.edu}

\affiliation{Department of Physics, University of California, Santa Barbara, CA 93106, USA}

\abstract{The concept of fixed-area states has proven useful for recent studies of quantum gravity, especially in connection with gravitational holography. We explore the Lorentz-signature spacetime geometry {\it intrinsic} to such fixed-area states in this paper. This contrasts with previous treatments which focused instead on Euclidean-signature saddles for path integrals that {\it prepare} such states. We analyze general features of fixed-area state geometries and construct explicit examples. The spacetime metrics are real at real times and have no conical singularities. With enough symmetry the classical metrics are in fact smooth, though more generally their curvatures feature power-law divergences along null congruences launched orthogonally from the fixed-area surface. While we argue that such divergences are not problematic at the classical level, quantum fields in fixed-area states feature stronger divergences. At the quantum level we thus expect fixed-area states to be well-defined only when the fixed-area surface is appropriately smeared.}

\begin{document} 

\maketitle

\section{Introduction} 
\label{sec:intro}

The study of entropy and quantum entanglement is a central focus of modern treatments of the AdS/CFT correspondence and its possible generalizations. In general, for a given boundary region $R$, the Hubeny-Rangamani-Takayanagi (HRT) \cite{Hubeny:2007xt} generalization of the Ryu-Takayanagi formula \cite{Ryu:2006ef} tells us that the entropy of region $R$ in the dual CFT is given by $\frac{A[\gamma_R]}{4G}$ where $G$ is the bulk Newton constant and $A[\gamma_R]$ is the area of the smallest extremal surface $\gamma_R$ satisfying both $\partial \gamma_R = \partial R$ and the requirement that $R$ and $\gamma_R$ be homologous within some Cauchy surface \cite{Headrick:2007km,Wall:2012uf}. The proof of this relation \cite{Dong:2016hjy} generalizes the Lewkowycz-Maldacena argument \cite{Lewkowycz:2013nqa} for the time-symmetric case.

As a result, the area $A[\gamma_R]$ of the HRT surface $\gamma_R$ plays a critical role in many discussions of AdS/CFT. It is thus natural to study bulk states in which the distribution of $A[\gamma_R]$ is sharply peaked with only very small fluctuations. Such `fixed-area' states were introduced in Refs.~\cite{Akers:2018fow,Dong:2018seb} to reproduce the entanglement properties of simple tensor network models of quantum error correction\footnote{Though one may also construct similar tensor network models with more general entanglement properties by adding additional degrees of freedom to the tensor network \cite{Donnelly:2016qqt}.}~\cite{Pastawski:2015qua,Hayden:2016cfa} and have since proved to be useful for a variety of constructions and analyses; see e.g. \cite{Bao:2018pvs,Dong:2019piw,Marolf:2019zoo,Penington:2019kki,Marolf:2020vsi,Dong:2020iod,Akers:2020pmf}. This is in part due to the fact that the replica trick is particularly straightforward to apply to fixed-area states, as there is a sense in which the usual back-reaction associated with replica numbers $n\neq 1$ vanishes for fixed-area states \cite{Akers:2018fow,Dong:2018seb}.

Our goal here is to explore and elucidate the spacetime geometries associated with such states. While the original works \cite{Akers:2018fow,Dong:2018seb} observed that saddles for Euclidean path integrals {\it preparing} such states will generally feature conical singularities at the fixed-area surface, the spacetime geometry {\it intrinsic to} fixed-area states has received relatively little attention. This has led to some confusion in the literature, especially with regard to the relation between fixed-area states and the microcanonical thermofield-double in the presence of a time-translation symmetry \cite{Goel:2020yxl}. We now discuss this apparent puzzle as an appetizer to our general treatment of the spacetime geometry of fixed-area states.

\subsection*{A possible confusion: the Microcanonical TFD vs Fixed-area states} 

\label{sub:puzzle}

\begin{figure}
    \centering
    \includegraphics[width=.8\textwidth]{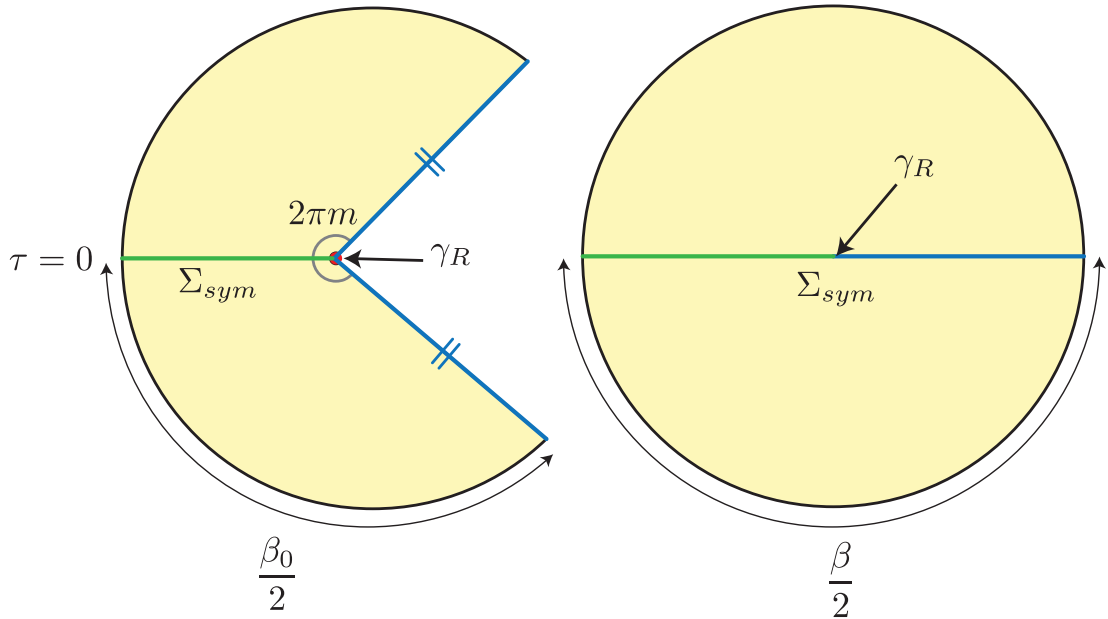}
    \caption{Left: Starting from a TFD state with inverse temperature $\beta_0$, fixing the area of the HRT surface $\gamma_R$, corresponding to the time slice $\tau=0$, results in a Euclidean saddle with a conical singularity (red) with opening angle $2\pi m$, where $m=\frac{\beta_0}{\beta}$. Right: The same state prepared as a microcanonical TFD state by imposing asymptotic fixed energy boundary conditions. This results in a smooth Euclidean saddle with boundary length $\beta$. The $\mathbb{Z}_2$ symmetric Cauchy slice $\Sigma_{sym}$ (green and blue) on both saddles has identical data and thus, results in identical Lorentzian spacetimes upon time evolution.}
    \label{fig:TFD}
\end{figure}

Fixed-area states may be constructed by starting from a seed state $\ket{\psi}$ and applying a quasi-projection operator that, for a given boundary region $R$,  restricts the probability distribution of the HRT area to be sharply peaked around a particular value $A_0$.   From Refs.~\cite{Akers:2018fow,Dong:2018seb}, it is also known that the entanglement spectrum of a fixed-area state is quite flat, so that the eigenvalues of the modular Hamiltonian $\hat{K}_R$ are also sharply peaked.

A particularly simple case is one in which the seed-state $\ket{\psi}$ is a thermofield-double (TFD) state for which the bulk geometry has a static Killing field with bifurcate horizon; i.e., in the bulk $|\psi\rangle $ describes a standard 2-sided black hole. Since $|\psi\rangle$ is the TFD state, the norm $\langle \psi |\psi \rangle$ is computed by a Euclidean thermal path integral with boundary $S^1 \times X$ for some $X$, where the metric on this boundary is also a product. It is easiest to visualize the associated bulk saddles when the bulk is 2-dimensional as in the case of Jackiw-Teitelboim gravity (where by `area' we mean the value of the dilaton). In that case, a bulk saddle can be represented as a disk in which there is a preferred point that represents the Euclidean horizon, see \figref{fig:TFD}.

We wish to consider the state constructed from $|\psi\rangle$ by fixing the area of the Euclidean horizon.   In this case, the Euclidean horizon coincides with the HRT surface for the region defined by taking all of $X$ at some point (which we may call $\tau=0$) on the $S^1$.   As described in Ref.~\cite{Dong:2018seb}, the corresponding fixed-area state is defined by the path integral with the same asymptotically AdS boundary conditions as the one that defines $|\psi\rangle$, but where the area of this HRT surface is also fixed to some $A_0$ as a boundary condition.  Since we do not integrate over that area, saddles for this path integral need not satisfy the corresponding equation of motion at the HRT-surface.  In particular, such saddles need not be smooth, and can instead have a conical singularity of arbitrary (constant) strength along the HRT surface.

If we consider saddles that preserve all symmetries, then in many cases there will be an analogue of Birkhoff's theorem which states that, at least locally, the possible bulk solutions are just the set of appropriately-symmetric (Euclidean) black holes. Fixing $A[\gamma_R]$ to some $A_0$ will then select precisely one such solution. But the period $\beta$ of the smooth Euclidean black hole with horizon area $A_0$ will not generally match the period $\beta_0$ of the $S^1$ at infinity. Nevertheless, we can  use the freedom to introduce a conical singularity at $\gamma_R$ (in this case with deficit angle $2\pi\left(1-\frac{\beta_0}{\beta}\right)$) to change the period of this solution to match $\beta_0$; see \figref{fig:TFD}.

On the other hand, as discussed in Ref.~\cite{Dong:2018seb}, one expects that in the leading semiclassical approximation the above fixed-area state will be equivalent to the microcanonical thermofield double so long as the area $A_0$ chosen above is not too small (so that the microcanonical ensemble is dominated by AdS-Schwarzschild black holes). The point here is that the seed state $|\psi\rangle$ above was chosen to be the usual (canonical) thermofield double, and so has modular Hamiltonian
\begin{equation}\label{eq:TFD}
	\hat{K}_{R} = \beta H + \log Z,
\end{equation}
where $H$ is the boundary Hamiltonian and the second term makes up the normalization. Furthermore, for each energy $E$ (again chosen to not be too small), the entropy is maximized by states that are well-described by an AdS-Schwarzschild black hole with horizon area $A$ determined by $E.$  As a result, restricting the canonical TFD to a narrow band of energies is essentially the same as restricting to a narrow band of areas.  

The interesting point then is that the restriction on energies can be implemented by performing an inverse Laplace transform.  This can be done semiclassically by integrating over boundary length $\beta$ to find a saddle with definite energy. Unlike the above fixed-area saddle, the corresponding Euclidean geometry is just a smooth disk with period $\beta$ at infinity determined as usual by the energy, or equivalently by the horizon area. See e.g. \cite{Marolf:2018ldl,Marolf:2022jra}.  Note also that in this case the period at infinity was {\it not} fixed as a boundary condition, but was determined dynamically by the saddle-point conditions.

Naively, it may appear that the conical defect appearing in the fixed-area state description might leave some singular imprint on the Lorentzian spacetime described by that state.  In contrast, it is clear that no such issue arises for the microcanonical TFD. Nevertheless, we will argue that these indeed lead to the same smooth Lorentzian classical solution.

Note in particular that the fixed-area Euclidean solution has a $\mathbb{Z}_2$ symmetry that leaves invariant a particular slice, $\Sigma_{sym}$. The $\mathbb{Z}_2$ symmetry implies that $\Sigma_{sym}$ has vanishing extrinsic curvature\footnote{Subtleties in this argument associated with the fact that $\Sigma_{sym}$ passes through the defect will be discussed in \secref{sec:SKsaddle} below.} $K_{ij}=0$, so the data on $\Sigma_{sym}$ also provides Cauchy data for a Lorentzian solution. Furthermore, the $U(1)$ symmetry of the Euclidean solution means that the induced metric $h_{ij}$ on $\Sigma_{sym}$ is just the usual induced metric on the surface of time-symmetry associated with the black hole of area $A_0$. In particular, the conical singularity leaves no imprint on either $h_{ij}$ or $K_{ij}$. Thus, the resulting Lorentzian spacetime is completely smooth
until one reaches the usual black hole singularities.  In particular, this Lorentzian spacetime is completely smooth at the HRT surface of interest.   And since the corresponding $\Sigma_{sym}$ in the microcanonical TFD saddle has precisely the same $h_{ij}$ and $K_{ij}$, it defines precisely the  same Lorentzian solution.

Now, the above setting is not generic, 
and it turns out that the Lorentzian solutions generally become singular when the U(1) symmetry is broken.  but these are power law singularities, not conical singularities or strict shockwaves.  We will describe the details of these singularities in \secref{sub:divergence_structure} below.


\subsection*{Overview} 

\label{sub:overview} 

Our treatment begins in \secref{sec:SK} with a brief review of fixed-area states and their preparation via path integrals. As mentioned above,   a common algorithm \cite{Akers:2018fow,Dong:2018seb,Dong:2019piw} for constructing fixed-area states involves first using a standard Euclidean (or, more generally, complex) path integral to construct a more familiar semiclassical bulk state and then modifying this prescription to fix the area of $A[\gamma_R]$. This then allows us to study the spacetime geometry of the fixed-area state in terms of the boundary conditions imposed on the above Euclidean path integral. \secref{sec:SK} extends previous such discussions by using Schwinger-Keldysh-like constructions to study the spacetime geometry intrinsic to the fixed-area state itself and to cleanly separate this geometry from that associated with sources used to prepare the state.

While fixed-area states can be of use in constructing  replica saddles, and while real-time replica saddles require complex metrics \cite{Colin-Ellerin:2020mva,Colin-Ellerin:2021jev}, we will show that the spacetime metrics in fixed-area states are generally real at real times. Furthermore, they have no conical singularities. 

As emphasized earlier, a $U(1)$ symmetric Euclidean solution results in a smooth Lorentzian spacetime. We thus  analyze various examples where this $U(1)$ symmetry is broken to demonstrate the features of generic fixed-area states. Our main analyses will be performed at the classical level, though we will comment on quantum effects at the end.

We first discuss a warmup example in \secref{sec:QFT} where we consider a non-gravitational scalar field theory in $1+1$ dimensions. This highlights the prominent features that we expect from fixed-area states such as the existence of power law divergences in the scalar field on the lightcone of the fixed-area surface.

We then move on to a general discussion of the structure of the Lorentzian spacetimes of fixed-area states in gravitational theories. We propose a general ansatz for the form of the classical solution in \secref{sec:gen-grav}. We then construct detailed examples in Jackiw-Teitelboim (JT) gravity and AdS$_3 \times X$ for compact $X$ in \secref{sub:jt_gravity_matter} and \secref{sub:ads_3} demonstrating the validity of our ansatz. The general structure of the above-mentioned singularities on the light cone of the fixed-area surface are analyzed in \secref{sub:divergence_structure}. We close with some final discussion in \secref{sec:disc} including comments on higher derivative and quantum corrections.


\section{Schwinger-Keldysh path integrals for fixed-area states} \label{sec:SK}

As described in \secref{sec:intro}, fixed-area states are simply states of gravitational systems in which the distribution of some HRT-area operator $A[\gamma_R]$ is sharply peaked, i.e. the width $\Delta A$ is small. Let us first discuss precisely what we mean by sharply peaked. As anticipated in Ref.~\cite{Bousso:2020yxi} and as established in Refs.~\cite{molly,xi}, in the semiclassical approximation the action of $\frac{A[\gamma_R]}{4G}$ is given by a so-called boundary-condition-preserving kink transform, which in particular induces a relative boost between the two entanglement wedges of some rapidity $s$.   From the uncertainty relation, we have
\begin{equation}\label{eq:unc}
	\Delta A \Delta s \gtrsim	O(G), 
\end{equation}
where $s$ is the relative boost between the two entanglement wedges on either side of the HRT surface. Depending on the value of $\Delta A$, we can classify fixed-area states into two types: pseudo-eigenstates and squeezed states. 

Pseudo-eigenstates are very sharply peaked and have $\Delta A \sim O(G^\alpha) $ with $\alpha \geq 1$. This leads to $\Delta s\rightarrow \infty$ in the semiclassical limit. As a result, we do not expect a single geometry to describe such states. On the other hand, squeezed states have $\Delta A \sim O(G^\alpha)$ with $\alpha \in (\frac{1}{2},1)$. Such states are expected to have a semiclassical description, and yet have $\Delta A$ parametrically smaller than states usually constructed by Euclidean path integrals \cite{Marolf:2018ldl}. Here we shall focus on such squeezed states and describe their associated Lorentzian spacetimes.

Now, fixed-area geometries must of course have a specified value of  $A[\gamma_R]$. But as noted above, so long as we consider the squeezed state case (so that $A[\gamma_R]$ is not specified too precisely), we expect that the state can remain semiclassical.  And since there are no other constraints, one further expects that all other aspects of the semiclassical spacetime geometry can be chosen arbitrarily (so long as they solve the equations of motion). However, as described in Refs.~\cite{Akers:2018fow,Dong:2018seb,Dong:2019piw}, one is typically interested in starting with some semiclassical bulk state $|\psi\rangle$, perhaps constructed using a gravitational path integral, and then applying a projection-like operator\footnote{We use the term ``projection-like operator'' to mean a Hermitian operator for which the variance of $A[\gamma_R]$ is small in the state $|\psi\rangle_{A_0} := \Pi_R|\psi\rangle$. We do {\it not} require $\Pi_R^2 = \Pi_R$. In particular, we might consider a Gaussian $\Pi_R = e^{-\frac{(A[\gamma_R]-A_0)^2}{2\sigma^2}}$ with some small width $\sigma$.} $\Pi_R$ that restricts this state to a range of $A[\gamma_R]$-eigenvalues of some small width $\Delta A$ about a central value $A_0$. We will thus investigate the spacetime geometries of fixed-area states that arise from such constructions and in particular their relation to the path integral boundary conditions used to define $|\psi\rangle$.

Recall that the squeezed state regime $\alpha\in(\frac{1}{2},1)$ described above suffices to fix the value of $A$ in the semiclassical limit $G\rightarrow 0$. In particular, in that limit we may use the recipe described in Refs.~\cite{Dong:2018seb,Dong:2019piw} for studying $|\psi\rangle_{A_0} := \Pi_R|\psi\rangle$. The recipe begins by supposing that we have already constructed a gravitational path integral that computes the original state $|\psi\rangle$, which in particular means that we are given boundary conditions for that path integral. From the bulk point of view we can think of the new state $|\psi\rangle_{A_0}$ as being created from $|\psi\rangle$ by the insertion of additional sources on $\gamma_R$, though of course the location of $\gamma_R$ must be determined dynamically in a manner that takes into account the back-reaction from those sources. As explained in Refs.~\cite{Dong:2018seb,Dong:2019piw}, in the saddle-point approximation this means that saddles for $|\psi\rangle_{A_0}$ can be taken to satisfy the same asymptotic AdS boundary conditions as $|\psi\rangle$ (with precisely the same sources at the asymptotic boundary), so long as we also 1) impose the usual equations of motion away from $\gamma_R$, 2) allow the bulk to have a codimension-2 conical singularity of arbitrary strength on a locus $\gamma_R$ homologous to $R$ and satisfying $\partial \gamma_R = \partial R$, 3) choose the strength of the conical singularity so that $A[\gamma_R] = A_0$, and 4) impose appropriate boundary conditions at $\gamma_R$.

In particular, in Euclidean signature, Appendix A of Ref.~\cite{Dong:2019piw} shows that the Euclidean Einstein-Hilbert action (including the delta-function term in the Ricci scalar associated with the conical singularity at $\gamma_R$) defines a good variational principle for this problem when the metric near $\gamma_R$ takes the following form:
\begin{gather}\label{eq:EBC} 
	ds^2 = dzd\bar z + T \frac{(\bar z dz-z d\bar z)^2}{z\bar z} + h_{ij} dy^i dy^j + 2iW_j dy^j (\bar z dz-z d\bar z),\\
	T=\hat{o}(1),\quad \pa_r T = \frac{\hat{o}(1)}{r}, \quad \pa_r h_{ij}=\frac{\hat{o}(1)}{r},\quad \pa_r W_j=\frac{\hat{o}(1)}{r},
	\label{eq:EBCdetails}
\end{gather}
where $z$ is defined as $z=r e^{im\theta}$ with $\theta\sim \theta+2\pi$, and $T$, $h_{ij}$, and $W_j$ are functions of all coordinates $(z,\bar z,y^i)$. Furthermore, $\hat{o}(1)$ denotes terms that vanish in the $r \rightarrow 0$ limit at least as fast as some power law $r^\eta$ with $\eta > 0$.\footnote{This fixes a typo in v1 of Ref.~\cite{Dong:2019piw}.} We refer to the conditions imposed by \Eqref{eq:EBC}, \eqref{eq:EBCdetails} as boundary conditions   to be imposed on Euclidean metrics at $\gamma_R$.  

We emphasize that the conical singularities on the surfaces $\gamma_R$ are associated with insertions of the operator $\Pi_R$ and, as such, they represent features of the way that the state $|\psi\rangle_{A_0}$ is being prepared rather than a feature intrinsic to the state itself. Indeed, if we can find another state $|\tilde \psi \rangle$ described by a smooth bulk saddle which yields the same fixed-area state up to quantum corrections
\begin{equation}
	|\tilde \psi \rangle_{A_0} = \Pi_R |\tilde \psi \rangle \approx \Pi_R |\psi \rangle = |\psi \rangle_{A_0}, 
\end{equation}
but where the saddle-point value of $A[\gamma_R]$ in the state $|\tilde \psi \rangle$ is already $A_0$, then the fixed-area saddle with asymptotically-AdS boundary conditions associated with the state $|\tilde \psi\rangle$ will be smooth regardless of the strength of the conical singularity in the original saddle defined by the asymptotically-AdS boundary conditions associated with $|\psi\rangle$.

To study the geometry intrinsic to the state $|\psi \rangle_{A_0}$, we should instead compute correlation functions in this state (which in the semiclassical limit should then factorize into a product of one-point functions). We thus consider 
\begin{equation}\label{eq:insertgs} 
	{}_{A_0}\langle \psi| g_{\mu_1\nu_1}(x_1)\dots g_{\mu_n\nu_n}(x_n) | \psi \rangle_{A_0} = \langle \psi| \Pi_R g_{\mu_1\nu_1}(x_1)\dots g_{\mu_n\nu_n}(x_n) \Pi_R | \psi \rangle, 
\end{equation}
where issues related to the gauge-dependence of the $g_{\mu_i \nu_i}(x_i)$ will not affect our discussion and will thus be ignored. Note that it is critical that there are two insertions of the operator $\Pi_R$ in \Eqref{eq:insertgs}. In particular, even if $\Pi_R$ were an exact projector, the fact that $\Pi_R$ will generally not commute with $g_{\mu_i \nu_i}(x_i)$ would make it difficult to use such a property to remove either copy of $\Pi_R$. Note also that the operators that sample the desired geometry naturally live {\it between} the two projectors.

We may thus construct a path integral that computes \Eqref{eq:insertgs} by first constructing path integrals for the bra and ket wavefunctions ${}_{A_0}\langle \psi| = \langle \psi| \Pi_R$ and $| \psi \rangle_{A_0} = \Pi_R | \psi \rangle$ and then using these wavefunctions as boundary conditions for a path integral that computes correlators of the $g_{\mu_i \nu_i}(x_i)$. The first step above is identical to constructing path integrals for the unconstrained bra and ket states $\langle\psi|$ and $|\psi\rangle$, except for the insertion of a constraint on the area of $\gamma_R$. Thus the final path integral involves constraints on two such surfaces $\gamma_R$ (though these surfaces may sometimes coincide).

We should of course add suitable sources $T^{\mu_i \nu_i}(x_i)$ to the action with respect to which we can vary to obtain the desired insertions of $g_{\mu_i \nu_i}(x_i)$. However, we see that in an appropriate sense we will make such variations only in the region between the two surfaces $\gamma_R$. Saddles for this problem will thus have two codimension-2 conical singularities, and it is only the region of the spacetime that in some sense lies between those singularities\footnote{We will explain the correct sense in more detail shortly. This sense will be clearest in Lorentz signature where the Cauchy problem is well-posed.} that can be directly interpreted as the geometry intrinsic to the fixed-area state $| \psi \rangle_{A_0}$. In particular, in the leading saddle-point approximation we can simply set the sources $T^{\mu_i \nu_i}(x_i)$ to zero and take the insertions of $g_{\mu_i \nu_i}(x_i)$ to report the saddle-point value of the metric at the point $x_i$. In that sense it is in fact sufficient to study the path integral that computes the norm 
\begin{equation}\label{eq:norm} 
	{}_{A_0}\langle \psi| \psi \rangle_{A_0} = \langle \psi| \Pi_R^2| \psi \rangle. 
\end{equation}

\subsection{Saddle points for fixed-area path integrals}
\label{sec:SKsaddle}

\begin{figure}
    \centering
    \includegraphics[width=.7\textwidth]{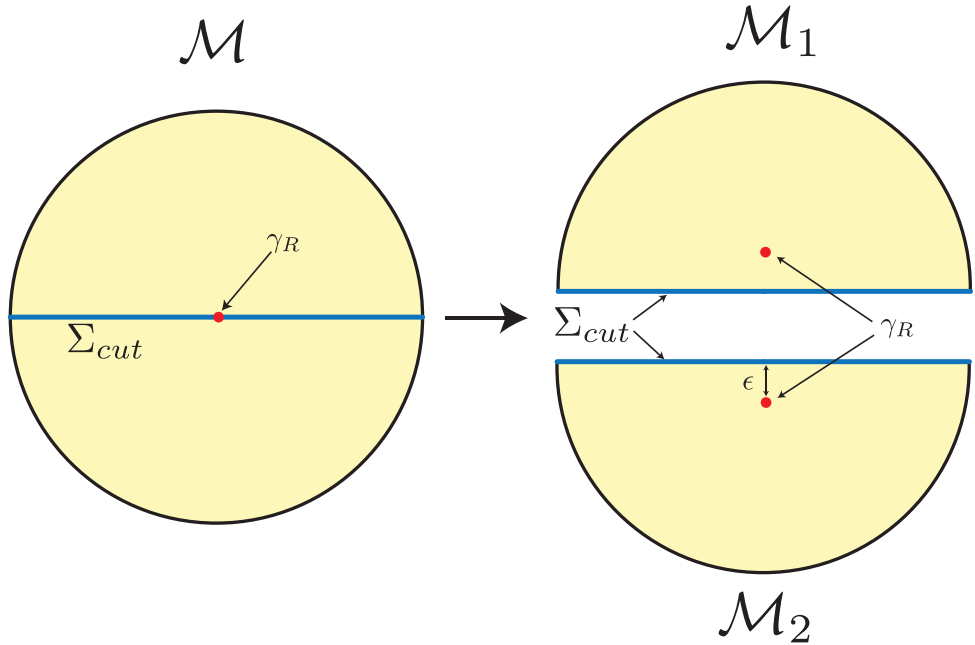}
    \caption{The saddle point manifold $\mathcal{M}$  for $\langle \psi| \Pi_R^2 | \psi \rangle$ can be split into two parts $\mathcal{M}_1$ and $\mathcal{M}_2$ along a slice $\Sigma_{cut}$ such that there are two conical defects (red) at $\gamma_R$, with one on the bra side of the cut and the other on the ket side. On the right, we have added a regulator $\epsilon$ that moves each of the resulting singularities away from the cut.  The original $\mathcal{M}$  should be understood as the limit $\epsilon\rightarrow 0$ where the two $\gamma_R$ surfaces coincide.}
    \label{fig:split}
\end{figure}

We now wish to describe the saddle points of this path integral. The constraints on the areas of the $\gamma_R$ surfaces mean that we do not integrate over these areas and, as a result, two of the Einstein equations need not be satisfied by our saddles, one at each of the two $\gamma_R$ surfaces. As explained in Refs.~\cite{Akers:2018fow,Dong:2018seb,Dong:2019piw}, this extra freedom allows conical singularities of arbitrary strength on each $\gamma_R$-surface. While the conical deficit or excess must be constant along each such surface, the fact that the constraints remove {\it two} Einstein equations means that the strengths of the singularities on the two $\gamma_R$ surfaces may be chosen independently. Furthermore, the idea that our path integral may be thought of as computing matrix elements associated with the bra and ket states ${}_{A_0}\langle \psi|$, $|\psi \rangle_{A_0}$, and that each contributes one of the conical singularities, suggests that one should be able to cut the saddle ${\cal M}$ into two pieces ${\cal M}_1, {\cal M}_2$ along some codimension-1 surface $\Sigma_{cut}$ (so that $\partial {\cal M}_1 \supset \Sigma_{cut} \subset \partial {\cal M}_2$) such that ${\cal M}_1, {\cal M}_2$ each contain only one of the $\gamma_R$ surfaces, see \figref{fig:split}. We will thus impose this requirement below, with the understanding that we think of each piece as being closed so that $\Sigma_{cut} \subset {\cal M}_{1}, {\cal M}_{2}$. Thus, this condition can be satisfied in the degenerate case where the two $\gamma_R$ surfaces at least coincide (in part or in whole) by taking $\Sigma_{cut}$ to pass through the locus of this coincidence. Note that this restriction requires the tangent spaces of the surfaces $\gamma_R$ to coincide at any point where the two surfaces intersect.

\begin{figure}
    \centering
    \includegraphics[width=.75\textwidth]{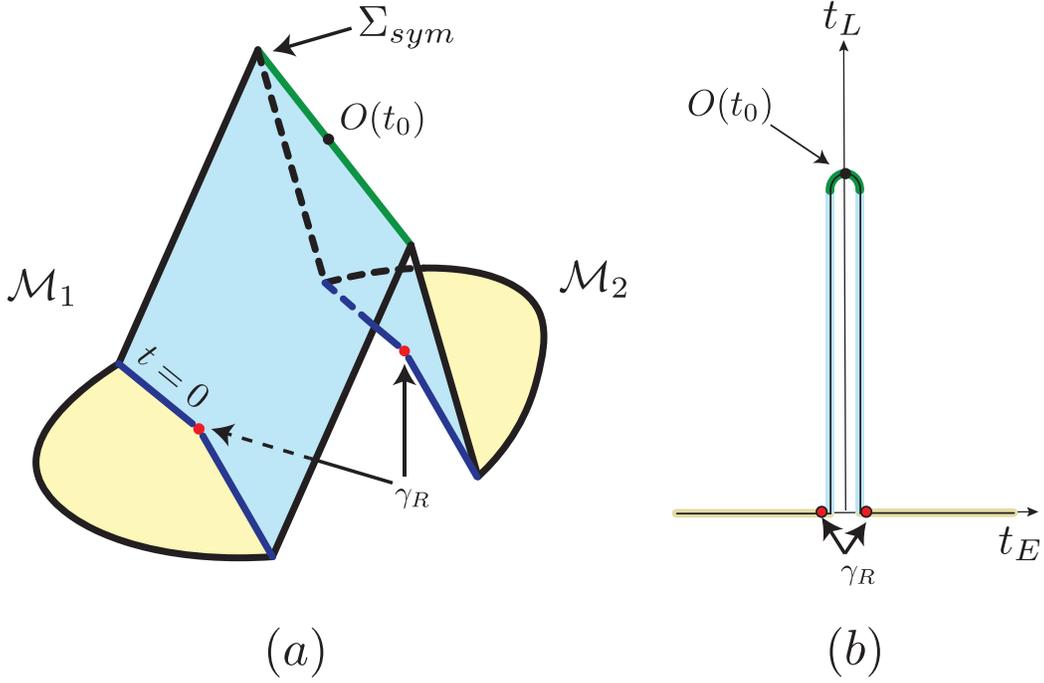}
    \caption{(a) The saddle point for $\langle \psi| \Pi_R (0) O(t_0) \Pi_R (0) | \psi \rangle$ has two conical defects (red) at $\gamma_R$ on the bra and ket side respectively. It has a $\mathbb{Z}_2$ symmetry that leaves invariant  the Cauchy slice $\Sigma_{sym}$ (green).  This slice splits the saddle into two parts $\mathcal{M}_1$ and $\mathcal{M}_2$. The region of the saddle point shaded light blue can be thought of as the spacetime inherent to the fixed-area state, while the yellow portion is involved in the preparation of the state.  We may take the blue portion to lie at real Lorentzian times and the yellow portion to lie at real Euclidean times.  In that sense, the two blue portions each involve the same interval of real Lorentzian times and should perhaps be drawn as being degenerate with each other, but we have separated the pieces for ease of visualization. (b) The corresponding Schwinger-Keldysh contour in the complex-time plane.}
    \label{fig:saddle}
\end{figure}

Now, in many cases the state $|\psi \rangle$ will have been constructed by specifying boundary conditions on a Euclidean asymptotically-AdS boundary. But we may nevertheless be interested in the metric $g_{\mu_i \nu_i}(x_i)$ at real times $t_i$. In this case our path integral will integrate over spacetimes which follow a Schwinger-Keldysh-like contour through the plane of complex times. Nevertheless, since $\Pi_R$ is Hermitian the expression \Eqref{eq:norm} is manifestly real. As a result, the path integral will have a $\mathbb{Z}_2$ symmetry that exchanges the parts of the boundary associated with boundary conditions for $\langle \psi|$ and for $|\psi \rangle$ and which simultaneously acts by complex-conjugation. Since this symmetry acts as a reflection (and conjugation) on the asymptotically AdS boundary, any bulk saddle ${\cal M}$ that preserves this conjugation symmetry must have a codimension-1 surface $\Sigma_{sym}$ that is invariant under the action of this $\mathbb{Z}_2$ symmetry\footnote{The argument uses the fact that our asymptotically AdS boundary conditions require ${\cal M}$ to have a $C^1$ conformal compactification so that at each point on the boundary the conformally rescaled spactime can be approximated by a Euclidean half-space. One may then show the existence of $\Sigma_{sym}$ by exhaustively studying the symmetries of the Euclidean plane.} as shown in \figref{fig:saddle}. We will consider only such saddles below.

For the moment, let us assume $\Sigma_{sym}$ does not intersect either surface $\gamma_R$ so that it lies in a smooth region of the saddle-point spacetime. In this case, $\Sigma_{sym}$ has a well-defined induced metric $h_{ij}$ and extrinsic curvature. We will use the symbol $K^E_{ij}$ to denote the extrinsic curvature defined using Euclidean conventions, and we will use $K^L_{ij} = i K^E_{ij}$ to denote the extrinsic curvature defined using Lorentz-signature conventions. Since all equations of motion are satisfied on $\Sigma_{sym}$, the data $h_{ij}, K^E_{ij}$ (or $h_{ij}, K^L_{ij}$), i.e., the metric and extrinsic curvature will clearly satisfy the relevant constraint equations. Furthermore, the invariance of $\Sigma_{sym}$ under the conjugation symmetry means that $h_{ij}$ must be real and the real part of $K^E_{ij}$ must vanish. This of course simply means that $K^E_{ij}$ is purely imaginary, or equivalently that $K^L_{ij}$ is real. In other words, just as in the analysis of real-time replica wormholes in Ref.~\cite{Colin-Ellerin:2020mva}, symmetry requires $\Sigma_{sym}$ to define Cauchy data appropriate to the Lorentz-signature initial value problem.

Let us now consider the case where $\Sigma_{sym}$ intersects some $\gamma_R$ at some point $x$. In this case one might worry that the extrinsic curvature of $\Sigma_{sym}$ at $x$ is not well-defined. In particular,  consider surfaces $\Sigma_\pm$ on either side of $\Sigma_{sym}$, where we take the conjugation symmetry to interchange $\Sigma_+$ and $\Sigma_-$. When there is a conical singularity at $x$, the surfaces $\Sigma_\pm$ will have different extrinsic curvatures at $x$ even in the limit where $\Sigma_\pm \rightarrow \Sigma_{sym}$. Indeed, with Euclidean signature conventions the real parts of their extrinsic curvatures will have delta-functions along $\gamma_R$ of opposite signs. In contrast, the conjugation symmetry requires that the imaginary parts of their $K^E_{ij}$ tensors match, so this imaginary part is continuous and unambiguous. 

Since the two surfaces $\gamma_R$ are in principle independent, the case where they intersect is just a degenerate limit of the more general case where they do not. As noted above, when they do not intersect the conjugation symmetry requires the real part of the extrinsic curvature of $\Sigma_{sym}$ to vanish. We will thus define this to also be the case when the $\gamma_R$ intersect. In effect, this is the statement that we should define the extrinsic curvature on $\Sigma_{sym}$ by first regulating the problem in a manner that separates the $\gamma_R$ surfaces (with $\Sigma_{sym}$ lying between them) as shown in \figref{fig:split}. We then take a limit where the regulator is removed.\footnote{Similar reasoning tells us that we can always take the conical singularities to lie inside the Euclidean region.  This means that the boundary conditions described by Eqs.~\eqref{eq:EBC} and \eqref{eq:EBCdetails} suffice to treat them.  The case where the conical singularities lies at the boundary between the Euclidean and Lorentzian region is to be regarded as a limit of the case where the singularities lie entirely in the Euclidean region of the contour.} But in practice it suffices to simply compute the (well-defined) imaginary part of $K^E_{ij}$ and to take the real part of $K^E_{ij}$ to vanish.

Now, when $\gamma_R$ intersects $\Sigma_{sym}$ at $x$, the conical singularity on $\gamma_R$ also means that some equations of motion are not satisfied at $x$. However, the fact that we required the existence of $\Sigma_{cut}$ means that the conjugation symmetry must exchange the two surfaces $\gamma_R$. As a result, each point of the fixed-point set $\Sigma_{sym}$ that intersects one copy of $\gamma_R$ must in fact intersect both copies. Furthermore, as noted in Ref.~\cite{Dong:2019piw} (see footnote 15), conical singularities can be thought of as arising from spacelike cosmic-brane sources that lie along $\gamma_R$. The effective stress tensor of such branes has non-zero components only tangent to $\gamma_R$, and in particular tangent to $\Sigma_{sym}$ at any point of intersection. But the constraints on initial data on $\Sigma_{sym}$ are associated with components of the equations of motion {\it normal} to $\Sigma_{sym}$, so they receive no contributions from such sources -- i.e., even when $\gamma_R$ intersects $\Sigma_{sym}$ the initial data on $\Sigma_{sym}$ satisfies the constraints that guarantee the initial value problem of the desired theory to be well-posed.

\begin{figure}
    \centering
    \includegraphics[width=.25\textwidth]{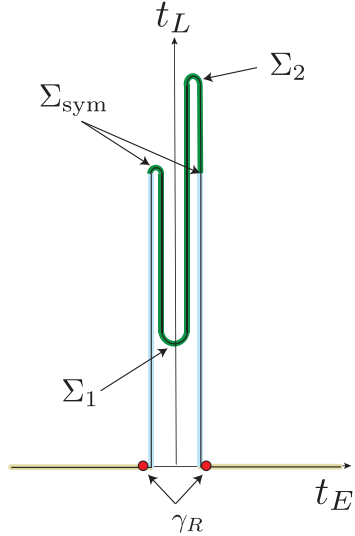}
    \caption{The saddle can be extended in the real time direction by evolving using the data on Cauchy slice $\Sigma_{\text{sym}}$. As an example, we depict the Schwinger-Keldysh contour for forward evolution up to surface $\Sigma_2$ and backward evolution up to surface $\Sigma_1$.}
    \label{fig:saddle_timefold}
\end{figure}

As a result, given any conjugation-symmetric saddle for the path integral that computes \Eqref{eq:norm}, we are free to extend it in the real time direction by evolving forward in time from $\Sigma_{sym}$ up to some new Cauchy surface $\Sigma_{2}$ and then backwards from $\Sigma_{2}$ back to $\Sigma_{sym}$. Indeed, we could just as well evolve backward in time from $\Sigma_{sym}$ to some $\Sigma_{1}$ and then forwards again from $\Sigma_1$ to $\Sigma_{sym}$ as shown in \figref{fig:saddle_timefold}, or we could even insert additional timefolds and evolve forwards and backwards in any manner that we like.  The new piece of this spacetime corresponding to our excursion along the real-time axis will be real and Lorentz signature, and will satisfy all of the equations of motion. In particular, it will be free of conical singularities. Futhermore, extending the saddle in this way leaves the action of the saddle unchanged because the factors of $e^{i S}$ associated with the forward-in-time parts of this evolution will precisely cancel the factors of $e^{-iS}$ associated with the backwards-in-time evolution. It is the geometry along this real-time excursion that we explore in more detail below. As a final remark we note that this part of the spacetime lies between the two surfaces $\gamma_R$ in the sense that it lies between two copies of the Cauchy surface $\Sigma_{sym}$, while the copies of $\gamma_R$ lie in the two regions between the copies of $\Sigma_{sym}$ and the Euclidean asymptotically AdS boundary. 

\section{Warmup: Scalar Field Theory in \texorpdfstring{$1+1$}{1+1} dimensions} \label{sec:QFT}

We would like to understand fixed-area states in the absence of a $U(1)$ symmetry. In order to do so, we first discuss a related-but-simpler problem involving a free scalar field $\phi$ in 1+1 dimensions on a fixed conical background. Due to the fixed background, this example cannot be directly interpreted as involving a fixed-area state.  Nevertheless, the analysis will highlight the key ideas needed for our discussion of dynamical gravity in \secref{sec:Gravity}.  There we will analyze the general structure of fixed-area states and illustrate it with various examples.

\begin{figure}
    \centering
    \includegraphics[width=.75\textwidth]{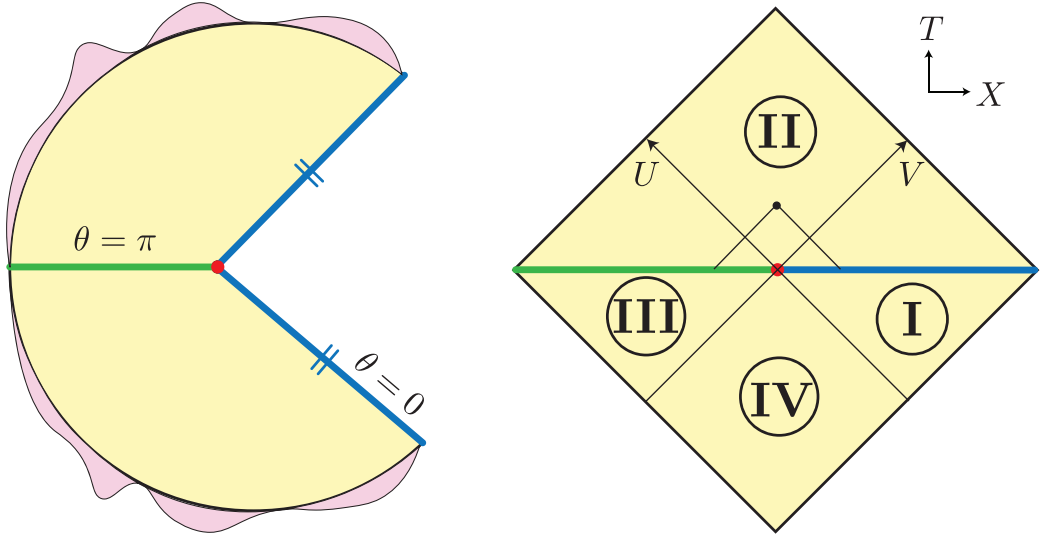}
    \caption{{\bf Left:} Euclidean preparation of the state of a scalar field on a conical background of opening angle $2\pi m$. A source for the scalar field (pink) is present at the asymptotic boundary and which breaks the $U(1)$ symmetry of the geometry.   This prepares initial data for Lorentzian evolution on the $\mathbb{Z}_2$ symmetric slice defined by the union of the blue half-line at $\theta=0$ line and the green half-line at  $\theta=\pi$.   Note that there is really only a single blue half-line; the marks on the diagram indicate that the two copies are to be identified.
    {\bf Right:} The initial data prepared by the Euclidean solution is used to generate the Lorentzian solution in Minkowski space by solving the equation of motion as described in the main text. The solution at the small black dot in region II is obtained by propagating the data along left and right-moving light rays (for a free massless field).}
    \label{fig:scalar}
\end{figure}

As discussed in \secref{sec:SK},  saddles for Euclidean path integrals preparing fixed-area states typically contain conical singularities.  Here we study a toy model of the influence of such singularities on solutions to equations of motion by considering saddles for a  scalar field path integral on a fixed conical background.  We can choose the time-contour of the background so that the saddle is purely Euclidean, or we can choose a contour for which the saddle contains pieces that evolve in real Lorentzian time.  

We take the background Euclidean geometry to be the simple $U(1)$-symmetric cone.  However, we will impose boundary conditions for the scalar field that break this $U(1)$ symmetry as shown in \figref{fig:scalar}.  In particular, the boundary conditions will preserve a $\mathbb{Z}_2$ conjugation symmetry.  Although the geometry in this example is flat, we will sometimes refer to these boundary conditions as `sources' using the terminology common in the AdS/CFT context.  

Any scalar field saddle 
on the Euclidean cone determines values of the scalar field $\phi$ on the $\mathbb{Z}_2$ symmetric Cauchy slice. Assuming that the saddle respects this $\mathbb{Z}_2$ symmetry, the real part of the normal derivative of $\phi$ must vanish.  

As a result, the data on this slice gives Cauchy data for a real solution to the equations of motion for our scalar in $1+1$-dimensional Minkowski spacetime (with no conical singularity). We will see that such solutions have power-law singularities on the past and future lightcones of the point $x=0$. We present the analysis for a massless scalar field in \secref{sec:massless} and for a massive scalar field in \secref{sec:massive}, both of which have qualitatively similar properties.

Since gravity is non-dynamical in this example, at some level we have simply chosen the Lorentzian spacetime by hand to be a singularity-free Minkowski space.  But one may view this choice as arising from a natural analogue of the discussion surrounding \figref{fig:saddle}.  If a regulator was introduced to split the original Euclidean conical singularity  (say, with conical deficit $\delta$) into two 
singularities (both with deficits $\delta/2$) while preserving $\mathbb{Z}_2$ symmetry, then the symmetric slice would have vanishing extrinsic curvature and induced metric $dx^2$.  This is precisely the data for the $t=0$ slice of Minkowski space.  One may also think of the Lorentz-signature Minkowski space as being generated by analytically-continuing the geometry from the Euclidean region between the singularities.

\subsection{Massless Field}\label{sec:massless}

Consider a free massless real scalar field on a fixed conical background. The Euclidean action is given by
\bal\label{massless-action}
S_{\rm massless} = \f{1}{2}\int d^2x \sqrt{g} (\pa \phi)^2,
\eal
with the standard metric given by
\bal
ds^2= dr^2+ m^2 r^2 d\theta^2, \qquad \theta \in [0,2\pi).
\eal
In this coordinate system, the time-reflection symmetric slice corresponds to $\theta=0$ and $\theta=\pi$ as shown in \figref{fig:scalar}.

In order to have a non-trivial classical solution, we turn on sources for the scalar field at some large distance cutoff boundary, $r=r_c$, and solve the equations for $r \le r_c$. A simple source we consider which breaks the $U(1)$ symmetry is the boundary condition
\bal\label{massless-source}
\phi_b(\theta) \equiv \phi(r_c ,\theta) = 2 r_c^{k/m} \cos (k \theta)
\eal
with some positive integer $k$. This source satisfies $\left. \pa_\theta \phi_b \right|_{\theta=0,\pi}=0$, as would any real source that preserves the $\mathbb{Z}_2$ symmetry $\theta \rightarrow -\theta$.  As a result, the slice defined by $\theta=0,\pi$ provides initial data for a real Lorentzian solution.  Solving the equation of motion associated with \Eqref{massless-action} is straightforward.  Furthermore, in order for our action to yield a well-defined variational principle for the scalar field, we require the solution to be finite at $r=0$. This uniquely determines the solution to be
\bal\label{massless-sol-euc}
\phi(r,\theta) = 2 r^{k/m} \cos(k\theta).
\eal
The desired Lorentzian initial data is found simply by setting $\theta=0,\pi$ in \Eqref{massless-sol-euc}. Defining the coordinate $X=r$ when $\theta=0$ and $X=-r$ when $\theta=\pi$, we can write the full initial value of $\phi$ in the form
\bal\label{massless-ini-dat}
\phi_0 (X) =  2 X^{k/m} \Theta(X) + 2 (-1)^k (-X)^{k/m}  \Theta(-X) .
\eal

When combined with the condition $\pa_T \phi(T,X)|_{T=0} = 0$, the data $\phi(T=0, X) = \phi_0(X)$ determines a unique Lorentzian solution on Minkowski space (which we take to have the standard metric $ds^2 = -dT^2 + dX^2$). 

While this example is simple enough that we could explicitly solve for $\phi(T,X)$ everywhere, it will be instructive to  first find the corresponding solutions in the left and right Rindler wedges marked as regions \rnum{1} and \rnum{3} in \figref{fig:scalar}. The point here is that, by causality, the analytic continuation $\theta \to  \theta+ i t$ of \Eqref{massless-sol-euc} is guaranteed to coincide  in regions \rnum{1} and \rnum{3} with the Lorentzian solutions constructed above.  This follows from the fact that the analytic continuation gives a Lorentzian solution with the correct initial data on both the positive and negative $X$-axes at $t=0$, together with the fact that such data defines a unique solution in each Rindler wedge\footnote{\label{foot:noan} Simple attempts to  extend the argument to include the origin $X=T=0$ will fail for the simple reason that the background Euclidean cone is not analytic at this point due to the delta-function Ricci scalar supported at the tip of the cone.  However, we describe a more sophisticated such extension in Appendix \ref{sec:analytic}.}. More explicitly, the above analytic continuation will give $\phi(X,T)$ in e.g.\ region \rnum{1} using $X=r \cosh( m t), T = r \sinh (m t)$. The solutions in regions \rnum{1}, \rnum{3} are thus given by
\bal\label{sol-massless-r1}
&\phi( U, V)  =   V^{k/m} + (-U)^{k/m} , &(\text{Region I}):\,V>0 \; {\rm and} \; U<0, \\
&\label{sol-massless-r3}\phi( U, V) =(-1)^k \l(  (-V)^{k/m} + U^{k/m} \r), &(\text{Region III}):V<0 \; {\rm and} \; U>0,
\eal
where $U,V$ are the null coordinates
\bal
V=T+X, \qquad U= T-X.
\eal

The key issue is then to find the solutions in regions \rnum{2} and \rnum{4}. Since the initial data is real, on general grounds we must find a real solution everywhere in the Lorentzian manifold. On the other hand, starting from \Eqref{sol-massless-r1} and \Eqref{sol-massless-r3}, one might naively expect to find a complex solution when $k/m$ is not an integer, as this is certainly the result of e.g. analytically continuing $V$ across the horizon at $V=0$. 

Luckily, the massless field is simple enough to allow us to clarify this issue by finding the explicit solutions in the past and future wedges. As is well-known, the most general solution to the massless $1+1$ Klein-Gordon equation is given by
\bal
\phi( U, V) = f(V) + g(U)
\eal
 for arbitrary functions $f, g$. Since the data in the left and right Rindler wedges fully determines both $f$ and $g$, the full solution must be
\bal\label{sol-massless-r2}
&\phi(U, V) =    V^{k/m} + (-1)^k (U)^{k/m} , &(\text{Region II}):\, V>0 \; {\rm and} \; U> 0, \\
&\label{sol-massless-r4}\phi(U, V) =   (-1)^{k} (-V)^{k/m} +  (-U)^{k/m} , &(\text{Region IV}):\,  V<0 \; {\rm and} \; U< 0.
\eal
Eqs. \eqref{sol-massless-r2} and \eqref{sol-massless-r4} are clearly real, solve the equation of motion, and induce the correct initial data on the surface $T=0$. Thus they give the desired solutions. We also describe an alternate construction of the same solution in Appendix~\ref{sec:massless-fourier} by using a plane-wave basis.

The important feature of Eqs.~\eqref{sol-massless-r2}, \eqref{sol-massless-r4} is that they display power law behaviour near the horizons.  Furthermore, for generic values of $m$, sufficiently high derivatives of $\phi(U,V)$ diverge. This can already be seen from the initial data at $T=0$ in \Eqref{massless-ini-dat}, which due to the time-symmetry of our solutions is closely related to the data on the horizons $U=0$ and $V=0$.  A similar feature will be true of the metric in the fixed-area states studied in \secref{sub:divergence_structure}, where it will lead to divergent tidal forces.

 \subsection{Massive Field}\label{sec:massive}
We now consider the case of a real massive scalar with a mass $\mu$ in a ``fixed-area" state in 1+1 dimensional Minkowski space. This example will play a key role in understanding higher dimensional examples in the presence of gravity. In particular, as we will see later, the equations near the horizon for a scalar field coupled to gravity in $AdS$ behave similar to the example studied below. Here the various coordinates and both the Euclidean and Lorentzian metrics are chosen to be the same as in \secref{sec:massless}.

The equation of motion for the massive field is given by
 \bal
 (\nabla^2 - \mu^2)\phi=0.
 \eal
 In Euclidean signature, we impose the following boundary conditions at $r=r_c$,
 \bal\label{eq:massive_bc}
 \phi(r_c , \theta) = 2 I_{k/m} (\mu r_c) \cos( k \theta).
 \eal
 The regular solution at the origin consistent with this boundary condition is 
\bal
\phi(r,\theta) =  2 I_{k/m} (\mu r) \cos(  k \theta).
\eal 
In much the same way as in the massless case, by changing coordinates to $(U,V)$  and performing an analytic continuation one may show the Lorentzian solution in the Rindler wedges to be given by
\bal
&\phi (U, V) = I_{k/m} (\mu \sqrt{-U V} ) \l((-V/U)^{\f{k}{2m}} + (-U/V)^{\f{k}{2m}} \r), &V \ge 0 , U \le 0, \nonumber\\
&\phi (U,V) =  (-1)^k  I_{k/m} (\mu \sqrt{-U V} ) \l((-V/U)^{\f{k}{2m}} + (-U/V)^{\f{k}{2m}} \r),&U \ge 0 , V \le 0,
\eal
where $I_{\nu}(x)$ is the modified Bessel function of the first kind. Note that as $\mu \to 0$, these solutions approach the massless solutions  \Eqref{sol-massless-r1} and \Eqref{sol-massless-r3} (up to a multiplicative constant associated with the different normalizations of \Eqref{massless-source} and \Eqref{eq:massive_bc}).

Given the solutions in the Rindler wedges, we can solve the equations of motion to extend the solution into regions \rnum{2} and \rnum{4}. In fact, this example is simple enough that we can simply guess the solution.  This is what we do below.   But the interested reader may consult Appendix~\ref{sec:massive_fourier} for a derivation of the result along the line $X=0$ performed by expanding the solution in terms of plane waves. 

If we guess that  solutions in regions \rnum{2} and \rnum{4} have the form  $(V/U)^{\f{k}{2m}} G( \sqrt{UV}) $, then the equation of motion sets the function $G$ to be a linear combination of Bessel functions $J_{k/m} (\mu \sqrt{ U V})$ and $Y_{k/m} (\mu \sqrt{ U V})$. One can check that only $J_{k/m} ( \mu \sqrt{U V})$ is consistent with continuity of the solution across the horizons (as $U, V \to 0$), and thus with the absence of delta-function sources in the equations of motion. The solution is thus given by
\bal
&\label{sol-massive-r2}\phi(U,V) =  J_{k/m}(\mu \sqrt{U V} ) \l( (V/U)^{\f{k}{2m}} + (-1)^k (U/V)^{\f{k}{2m}}\r) \qquad U \ge 0, V \ge 0, \\
&\label{sol-massive-r4}\phi(U,V) =  J_{k/m}(\mu \sqrt{U V} ) \l( (U/V)^{\f{k}{2m}} + (-1)^k (V/U)^{\f{k}{2m}}\r) \qquad U \le 0, V \le 0.
\eal

Note that in (for example) region \rnum{2}, the limit $ U V \to 0$ gives $\phi(U,V) \sim  V^{k/m} + (-1)^k U^{k/m}$ which coincides with \Eqref{sol-massless-r2}. So for the types of sources we have considered, the leading behaviour of the solution near the lightcone is determined by the massless solution. In particular, the power law divergences found in \secref{sec:massless} are not tied to the massless nature of that example, but are generic for all masses.  We will find a similar feature to be true in arbitrary theories of gravity coupled to matter.

\section{Fixed-area states in gravity} \label{sec:Gravity} 

We now turn our attention to the Lorentzian geometry of fixed-area states in the presence of dynamical gravity. We will show how to obtain the Lorentzian metric from the Euclidean path integral which prepares the fixed-area state. In \secref{sec:gen-grav}, we begin with a discussion of the general structure of fields as power series expansions near the HRT surface in both Euclidean and Lorentzian signature. Then, we illustrate the prescription for constructing the Lorentz-signature fixed-area state solutions in two simple examples. In \secref{sub:jt_gravity_matter}, we construct exact fixed-area solutions in JT gravity coupled to a scalar field. In \secref{sub:ads_3}, we construct fixed-area solutions in AdS$_3$ gravity coupled to a scalar field by including the effects of backreaction perturbatively. This example illustrates the generic features of fixed-area states in higher dimensional gravity since it arises from dimensional reduction. Thus, it complements the example in JT gravity. Finally,  \secref{sub:divergence_structure} discusses singularities on the horizons of fixed-area states in the presence of arbitrary boundary sources.

\subsection{General Lorentzian solutions} \label{sec:gen-grav} 

Before discussing the structure of the Lorentzian solutions, let us first review the structure of Euclidean conical solutions as analyzed in Ref.~\cite{Dong:2019piw}. The metric near the codimension-2 defect at the HRT surface $\gamma_R$ takes the form 
\bal\label{near-defect-euc}
ds^2&= dz d\bar{z} + T \f{(\bar{z} dz - z d\bar{z})^2}{z \bar{z}} + h_{ij} dy^i dy^j + 2 i W_j dy^j (\bar{z} dz - z d\bar{z}),
\eal
where $y^i$ denotes the transverse directions and $(z,\bar{z})$ represent the normal directions to $\gamma_R$.  Note $z$ may be written as $z=r e^{im\theta}$ with $\theta\sim \theta+2\pi$. The quantities $T, h_{ij}$ and $W_j$ are functions of all the coordinates $(z,\bar{z}, y^i)$. For Einstein gravity minimally coupled to scalar matter with standard two-derivative actions, the metric components in general have the following power series expansion in terms of powers of $z,\bar{z}$ in a neighbourhood of $\gamma_R$:
\bal
T&= \sum_{\substack{p,q,s=0 \\ pq>0 \; \text{or}\; s>0} }^\infty T_{pqs}(y^i) z^{p/m} \bar{z}^{q/m} (z \bar{z})^s, \label{T-function} \\
W_i&= \sum_{\substack{p,q,s=0 } }^{\infty} W_{i,pqs}(y^i) z^{p/m} \bar{z}^{q/m} (z \bar{z})^s,\label{U-function}\\
h_{ij}&= \sum_{\substack{p,q,s=0 } }^{\infty} h_{ij, pqs}(y^i) z^{p/m} \bar{z}^{q/m} (z \bar{z})^s\label{h-function},
\eal
where $2\pi m$ is the opening angle of the conical defect at $\gamma_R$. Any scalar matter field $\phi$ has a similar series expansion near the conical defect:
\bal \label{eq:matter}
\phi &= \sum_{p,q,s=0}^{\infty} \phi_{pqs}(y^i) z^{p/m} \bar{z}^{q/m} (z \bar{z})^s.
\eal
In particular, it was demonstrated in Ref.~\cite{Dong:2019piw} that such a power series expansion near $\gamma_R$ provides a good variational ansatz for finding Euclidean conical solutions with specified asymptotic boundary conditions. 

\begin{figure}
    \centering
    \includegraphics[width=.45\textwidth]{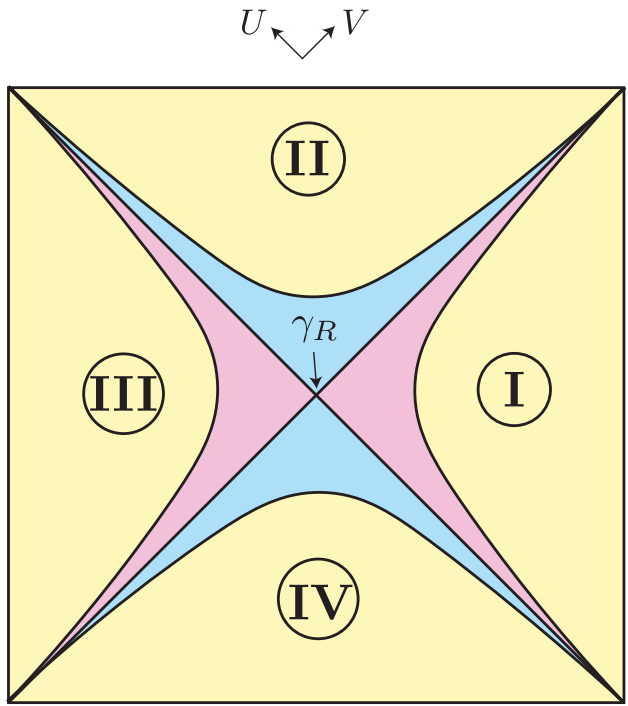}
    \caption{The Lorentzian spacetime can be divided into four wedges with respect to the HRT surface $\gamma_R$ labelled I-IV. The power series expansion we propose is valid in the regions shaded pink and blue. The solution in the pink region is obtained by analytically continuing the Euclidean solution. The solution in the blue region is obtained by a simple transformation of the analytically continued solution.}
    \label{fig:power}
\end{figure}

As discussed previously, the Euclidean solution can be analytically continued to obtain a solution in each of the Rindler wedges defined by the HRT surface $\gamma_R$. We promote the complex coordinates $z,\z$ to lightcone coordinates $U,V$ and assuming that the power series expansion is valid in a tubular neighbourhood $|z \z|\leq r_0^2$ around $\gamma_R$, we obtain a Lorentzian solution in a corresponding hyperboloidal region $-r_0^2 \leq UV < 0 $ as shown in \figref{fig:power} (shaded pink). For any field $\phi(U,V)$, we obtain a power-series expansion in the Rindler wedges of the form
\bal\label{eq:rindler}
&\phi (U,V) = \sum_{p,q,s} \phi_{pqs} V^{p/m} (-U)^{q/m} (-UV)^s, &U \le 0 , V \ge 0, \nonumber\\
&\phi (U,V) =  \sum_{p,q,s} (-1)^{p+q} \phi_{pqs} (-V)^{p/m} U^{q/m} (-UV)^s,&U \ge 0 , V \le 0,
\eal

In order to construct the solution in the future and past wedges, we first demonstrate a simple procedure for generating new solutions to the equations of motion. Given a solution for a field $\phi(U,V)$ in terms of the coefficients $\phi_{pqs}$, another solution is given by the set of coefficients $\tilde{\phi}_{pqs}=e^{i \alpha_1 p} e^{i \alpha_2 q} \phi_{pqs}$, where $\alpha_{1,2}$ are arbitrary constants. This can be easily seen in the case when $m$ is chosen to be irrational, since terms of the form $V^{p/m} U^{q/m}$ are linearly independent for different values of $p,q$. In this case, at any given order the equations of motion generate a relation for the coefficient $\phi_{pqs}$ in terms of a combination of homogenous product of coefficients such as $\phi_{p_1 q_1 s_1} \phi_{p_2 q_2 s_2}$ with $p_1+p_2=p$ and $q_1+q_2=q$. This makes it clear that multiplying the coefficients by a phase that linearly depends on $p$ or $q$ doesn't alter the defining relation arising from the equation of motion. Thus, the new set of coefficients $\tilde{\phi}_{pqs}$ also solves the equation of motion.

For rational values of $m$, one can take the limit from irrational $m$ and the limit is well-behaved as demonstrated in Ref.~\cite{Dong:2019piw}. Since the Lorentzian equations of motion we are analyzing here are Wick-rotated versions of the Euclidean equations of motion analyzed there, the same arguments go through in our case.

In general, this new solution would not be as useful since it is complex while one is usually interested in a real solution given real initial data. However, in our problem, this transformation will turn out to be useful in generating real solutions in the past and future wedges. In particular, we propose that the solution in these regions take the form:
\bal \label{eq:past}
&\phi (U,V) = \sum_{p,q,s} (-1)^{s+q} \phi_{pqs} V^{p/m} U^{q/m} (UV)^s, &U \ge 0 , V \ge 0,\nonumber\\
&\phi (U,V) =  \sum_{p,q,s} (-1)^{p+s} \phi_{pqs} (-V)^{p/m} (-U)^{q/m} (UV)^s,&U \le 0 , V \le 0,
\eal
The above power series can be obtained from \Eqref{eq:rindler} by applying a phase rotation to the coefficients that is linear in $p,q$ as described above.  This ensures that it solves the equations of motion. It is also easy to check that this ansatz agrees with 
\Eqref{eq:rindler} on all horizons, and so satisfies the desired initial data on these null surfaces.  Uniqueness of the null initial value problem then guarantees that the correct solution will be of this form.  Indeed,  one can directly check that the solutions in \secref{sec:QFT} take this form. We now demonstrate this procedure in more detail in various explicit examples in JT gravity and $AdS_3$.

\subsection{Example 1: JT Gravity + Matter} 
\label{sub:jt_gravity_matter}
We begin with an example involving JT gravity minimally coupled to a massless real scalar field. We use $\Phi$ to represent the dilaton in order to distinguish it from the matter field $\phi$. This case is sufficiently simple that the equations of motion can be solved exactly  with arbitrary boundary sources \cite{Almheiri:2014cka}. We first solve for the dilaton profile in Euclidean signature, and then obtain the Lorentzian solution using the initial value formulation.

The Euclidean metric for a fixed-area state in AdS$_2$ is
\bal
ds^2= \frac{dz d\zb}{(1-z \zb/4)^2},
\eal
where $z= r e^{ i m \theta}$ and $\bar z=r e^{- i m \theta}$ with $m$ chosen to satisfy boundary conditions at Euclidean infinity. On the cutoff surface $z\zb=r_c^2=4(1-m \epsilon)^2$ where $\epsilon$ is small, we impose a boundary condition on the matter field $\phi$,
\bal
\phi=2 r_c^{k/m} \cos(k\theta).
\eal
The equation of motion for the massless scalar field is invariant under conformal transformations and thus, the solution is the same as the solution on a flat cone in \secref{sec:massless}. Thus, we obtain the solution
\bal
\phi = z^{k/m} + \bar{z}^{k/m}.
\eal
The stress tensor for this solution is
\bal
T_{zz} (z) = (k/m)^2 z^{2(k/m-1)},\quad T_{\zb \zb} (\zb) = (k/m)^2 \zb^{2(k/m-1)},\quad T_{z\zb} =0,
\eal
and the dilaton equation of motion is \cite{Almheiri:2014cka}
\bal
- (1-z \zb/4)^{-2} \partial_{\mu} \left((1-z \zb/4)^{2}\partial_{\nu}\Phi\right)=T_{\mu \nu}.
\eal
We thus find 
\bal
\label{eq:EuclideanDilaton}
\Phi &=\int^z dx \f{(1- \zb x/4) (x-z)}{(1-z\zb/4)} T_{zz}(x) +  \int^{\zb}  dx \f{(1- z x/4) (x-\zb)}{(1-z\zb/4)} T_{\zb \zb}(x) +\Phi_0 \nonumber \\
&= (z^{2k/m} + \zb^{2k/m}) \f{(k/m)  \left((1+2k/m)+ (1-2k/m) z \zb/4) \right) }{2(1-4k^2/m^2)(1-z \zb/4)}+\Phi_0,
\eal
where $\Phi_0$ can be any solution to pure JT gravity without matter, i.e.,
\bal
\Phi_0=\frac{a+dz\zb}{1-z\zb},
\eal
for any constants $a$ and $d$.
Note that the ambiguity in choosing the lower limits of the integrals can be absorbed into $\Phi_0$, and possible linear terms of the form $b z+c\zb$ in the numerator of $\Phi_0$ are discarded since they do not respect the $\mathbb{Z}_2$ symmetry and periodicity $\theta \to \theta+2\pi$. 

We choose to study an example where the pure gravity solution is the fixed-area state defined by a Euclidean black hole at inverse temperature $\beta$, so that \cite{Almheiri:2014cka}
\bal
\Phi_0 = m \Phi_b \f{1 +  z \zb/4}{1- z \zb/4},
\eal
with $\beta/\epsilon$ the length of the boundary curve and $\Phi_b/\epsilon$ the boundary dilaton value.\footnote{One may observe that, in our full solution \eqref{eq:EuclideanDilaton}, as we approach the cutoff surface $z\zb \to r_c$, the dilaton $\Phi$ does not approach a constant.  This may appear inconsistent with the usual boundary condition $\Phi|_{r=r_c} = const$ in JT gravity. However, we can instead choose to reformulate the boundary conditions by finding a surface on which the dilaton is in fact constant in our solution and then using the scalar field profile on the given surface as the boundary condition we impose for the scalar field.  We simply choose the above formulation for convenience.}

The next step is to inspect the initial value for the dilaton on the $T=0$ surface, which will then be used to construct the Lorentzian solution. The $\mathbb{Z}_2$ symmetric slice is given by the union of the surfaces $\theta=0$ and $\theta=\pi$. For $0<X<2$ we take $X=r$, and for $-2<X<0$ we take $X=-r$. The initial value for $\Phi$ is thus
\bal
\label{eq:InitialDilaton}
\nonumber \Phi|_{T=0}(X)=&\left( X^{2k/m} \Theta(X)+(-X)^{2k/m} \Theta(-X) \right) \f{(k/m)\left((1+2k/m) +(1-2k/m)  X^2/4\right) }{(1-4k^2/m^2)(1-X^2/4) } \\
&+m \Phi_b \frac{1+ X^2/4}{1-X^2/4}.
\eal
In the Lorentzian spacetime, the metric may always be written in the form
\bal
ds^2= -\frac{dU dV}{(1+U V/4)^2},
\eal
with the asymptotic boundaries being located at $UV=-4$.

We start by finding the solution for the scalar field. Since this field satisfies the massless wave equation, and since this wave equation is conformally invariant, the result is identical to that in the non-gravitational example in \secref{sec:massless}, i.e.,
\bal
\phi(U,V)&=V^{k / m}+(-U)^{k / m},& \text { (Region I) }: V>0 \text { and } U<0, \\
\phi(U, V) & =V^{k / m}+(-1)^{k}(U)^{k / m}, & \text { (Region II) }: V>0 \text { and } U>0, \\
\phi(U, V)&=(-1)^{k}\left((-V)^{k / m}+U^{k / m}\right), & \text { (Region III) }: V<0 \text { and } U>0,\\
\phi(U, V) & =(-1)^{k}(-V)^{k / m}+(-U)^{k / m}, & \text { (Region IV) }: V<0 \text { and } U<0.
\eal
This yields the stress tensor
\bal
T_{UU}(U)&=(k/m)^2\left(U^{2(k/m-1)} \Theta(U)+(-U)^{2(k/m-1)} \Theta(-U) \right),\\
T_{VV}(V)&=(k/m)^2 \left(  V^{2(k/m-1)} \Theta(V)+(-V)^{2(k/m-1)} \Theta(-V) \right),\\
T_{UV}&=0.
\eal
Following Ref.~\cite{Almheiri:2014cka}, the dilaton solution in the Lorentzian spacetime can then be computed as 
\bal
\label{eq:LorentzDilaton}
\Phi= \int^{U} dx \f{(1+ V x/4) (U -x)}{(1+UV/4)} T_{UU}(x) +  {\int^{V}} dx \f{(1+ U x/4) (V - x)}{(1+UV/4)} T_{VV}(x) +\tilde \Phi_0,
\eal
where 
\bal
\tilde \Phi_0= \frac{A+B U+C V+D UV}{1+UV/4}.
\eal

The full Lorentzian solution can then be constructed by performing the integrals in \Eqref{eq:LorentzDilaton} and choosing integration constants to the prescribed initial values from \Eqref{eq:InitialDilaton}. Doing so leads us to the result
\bal
\Phi=f(U,V)\f{(k/m)  \left((1+2k/m)- (1-2k/m) UV/4 \right) }{2(1-4k^2/m^2)(1+UV/4)}+m \Phi_b \frac{1- UV/4}{1+UV/4},
\eal
where
\bal
f(U,V)&= V^{2k/m} + (-U)^{2k/m}, \quad &\text{(Region I):}~ U<0,V>0\\
f(U,V)&= V^{2k/m} + U^{2k/m}, \quad &\text{(Region II):}~ U>0,V>0\\
f(U,V)&=(-V)^{2k/m} + U^{2k/m}, \quad &\text{(Region III):}~ U>0,V<0\\
f(U,V)&=(-V)^{2k/m} + (-U)^{2k/m}, \quad &\text{(Region IV):}~ U<0,V<0.
\eal
One can then verify that the above expressions solve the equations of motion, and one can also check that the solution satisfies the general power series expansion described in \secref{sec:gen-grav}. Importantly, despite the appearance of singularities on the lightcone of the HRT surface, the solution is smooth in the interior of both the past and future wedges.


\subsection{Example 2: \texorpdfstring{$AdS_3$}{AdS3} coupled to a massless scalar field}
\label{sub:ads_3}

We now consider an example of a fixed-area state geometry for $AdS_3$ gravity minimally coupled to a real scalar field. The corresponding Euclidean problem was analyzed in detail in  Ref.~\cite{Lewkowycz:2013nqa} for the case $m=1/n$ with $n \in \mathbb{Z^+}$ where it was interpreted as a Euclidean replica geometry. Here we extend the calculation in Ref.~\cite{Lewkowycz:2013nqa} to conical defects with arbitrary strength $m$ and focus on the Lorentzian solution whose initial data agrees with the Euclidean solution on the surface of time-symmetry.

An interesting feature of this example is the connection to gravity in higher dimensions. As mentioned in Ref.~\cite{Lewkowycz:2013nqa}, solutions of the form $AdS_3 \times S^3 \times T^4$ give rise to three dimensional gravity coupled to two massless real scalar fields. One obtains scalar fields $\phi_1$ and $\phi_2$ from the metric components of $T^4$ when parametrized in the form
\bal
ds^2_{T^4} = e^{2\phi_1} dy_1^2 + e^{2\phi_2} dy_2^2+ e^{-2\phi_1} dy_3^2 + e^{-2\phi_2} dy_4^2.\eal
Studying the fixed-area state solution of the scalar field and metric components in this example is thus directly related to fixed-area states in pure gravity in higher dimensions. In particular, as we will discuss, the Lorentz-signature singularities of the scalar field that arise from the presence of the conical defect in the Euclidean preparation correspond to geometric singularities (and generally to Weyl curvature singularities) of the higher-dimensional gravity theories. For simplicity and clarity, we will set $\phi_2=0$ and denote $\phi=\phi_1$ as the only relevant scalar field.

We now turn to the actual preparation of our fixed-area state solution.
As usual, we start in Euclidean signature and use the result to determine the initial data for the Lorentzian scalar field and metric. Following Ref.~\cite{Lewkowycz:2013nqa}, we solve the Euclidean equation of motion for the massless real scalar field and the metric perturbatively by assuming that the source for the scalar field is small. To zeroth order, the metric of Euclidean $AdS_3$ with a conical defect of strength $m$ is given by
\bal \label{eq:ads3}
ds^2 = \frac{dr^2}{r^2+ 1} + m^2 r^2 d\theta^2 + m^2(r^2+ 1) dy^2,
\eal
where $\theta \in [0, 2\pi)$. Here we set $l_{AdS}=1$. The transverse coordinate $y$ does not play a role for the class of solutions considered here and can be chosen to be either compact or non-compact. The factor of $m^2$ in the last term in \Eqref{eq:ads3} is chosen to ensure that when $y$ is compact with fixed period, the conformal structure of the boundary torus is independent of $m$. 

We turn on a non-trivial source for the scalar field by imposing the following boundary condition at a cutoff surface $r=r_c$,
\bal
\l. \phi \r|_{r=r_c} = 2 \eta \cos( k \theta).
\eal
The equation we are solving is
\bal
\nabla_\mu \nabla^\mu \phi = 0,
\eal
and the solution is then given by
\bal \label{scalar-ads3-sol}
\phi = 2 \eta \cos(k \theta) \f{f_{m,k}(r)}{f_{m,k}(r_c)}, \qquad f_{m,k}(r) = r^{k/m} {_2F_1} \l( \f{k}{2m} , \f{k}{2m}+1 , \f{k}{m}+1, -r^2\r).
\eal
Using the solution for the scalar field, the stress tensor can be decomposed in terms of Fourier modes,
\bal
T_{\mu \nu} &= \nabla_\mu \phi \nabla_{\nu} \phi - \f{g_{\mu\nu}}{2}  (\nabla \phi)^2\\
&=T^0_{\mu\nu} +  T_{\mu\nu}^+ e^{2ik\theta} + T_{\mu\nu}^- e^{-2ik\theta}.
\eal
 We can then use the above result to solve the metric equations at leading order in $\eta$. Following the ansatz in Ref.~\cite{Lewkowycz:2013nqa}, we write the perturbed metric as
\bal\label{eq:ansatz}
ds^2 = \f{dr^2}{r^2+1 + g(r,\theta)} + m^2 r^2 d\theta^2  + m^2(r^2+1) (1+v(r,\theta)) dy^2,
\eal
where $g(r,\theta),v(r,\theta)$ consist of three fourier modes in the angular direction, e.g.,
\bal
g(r,\theta)= g_0(r)+g^{+}(r) e^{2 i k \theta}+g^{-}(r) e^{-2 i k \theta}.
\eal
The Einstein equations after dimensional reduction are then given by,
\bal
R_{\mu \nu} - \f{g_{\mu \nu}}{2} (R+2) = 8 \pi G T_{\mu \nu}, \qquad T_{\mu \nu } = \nabla_\mu \phi \nabla_{\nu} \phi - \f{g_{\mu\nu}}{2}  (\nabla \phi)^2. 
\eal
By using the ansatz in \Eqref{eq:ansatz}, one can find integral expressions for $g(r,\theta)$ and $v(r,\theta)$. However, the closed form solutions for $v(r,\theta)$ or $g(r,\theta)$ will not be important for the discussion of the Lorentzian evolution. The interested reader may refer to Appendix~\ref{sec:ads3_app} for more details.

So far, we have considered the equations and solution in Euclidean signature. In order to write down the Lorentzian solution, we follow identical steps to \secref{sec:QFT}. Let us first discuss the undeformed metric in the Lorentzian signature. The initial slice is given by the union of the slices $\theta=0$ and $\theta=\pi$. By continuing $\theta \to \theta + i t$, on the $\theta=0, \pi$ slices we find
\bal\label{emptads3}
ds_L^2 = -m^2 r^2 dt^2 + \f{dr^2}{r^2+1} + m^2 (r^2+1) dy^2.
\eal
By rescaling the coordinates, the metric can be written in the more suggestive form
\bal\label{ads3-rescaledu1}
ds_L^2 =  - r^2 {d\tilde{t}}^2 + \f{dr^2}{r^2+1} +  (r^2+1) d\tilde{y}^2, \qquad \tilde{t} = m t, \qquad \tilde{y} = m y,
\eal
which is just the familiar Lorentz signature $AdS_3$ metric with a rescaled transverse coordinate $y \to m y$. In particular, the geometry in \Eqref{ads3-rescaledu1} is manifestly smooth in Lorentz signature.

In order to make the Lorentzian solution more explicit, let us write the undeformed metric \eqref{emptads3} in global coordinates
\bal
ds_L^2 = -\f{4 dU dV}{(1+  U V )^2} + m^2 \l(\f{1- U V}{1+ U V} \r)^2 dy^2,
\eal
where 
\bal
U= -\f{r}{1+\sqrt{1+r^2}} e^{ - m t}, \; V= \f{r}{1+\sqrt{1+r^2}} e^{ m t},
\eal
are the global null coordinates. As in the discussion of \secref{sec:QFT}, the Lorentzian solution for the scalar field inside the ``Rindler" wedges is straightforward to obtain by analytically continuing \Eqref{scalar-ads3-sol} to find 
\bal\label{ads3-scalar-rindler}
&\phi= \frac{\eta}{f_{m,k}(r_c)}\; \f{(-2U)^{k/m} + (2 V)^{k/m}} {\l(1+ U V\r)^{k/m}}  {_2F_1} \l(\f{k}{2m}, \f{k}{2m}+1, \f{k}{m}+1, \f{4 U V}{(1+ U V)^2} \r),&U<0,\; V>0,\nonumber\\
& \phi= \frac{\eta}{f_{m,k}(r_c)}\; (-1)^k \f{(2U)^{k/m} + (-2 V)^{k/m}} {\l(1+ U V\r)^{k/m}}  {_2F_1} \l(\f{k}{2m}, \f{k}{2m}+1, \f{k}{m}+1, \f{4 U V}{(1+ U V)^2} \r),&U>0,\; V<0.
\eal

We can now obtain the solution in the past and future wedges by solving the equations of motion and ensuring continuity of the fields across the horizons.  Doing so yields 
\bal
&\phi=  \frac{\eta}{f_{m,k}(r_c)}\;  \f{(-2U)^{k/m} + (-1)^k (-2 V)^{k/m}} {\l(1+ U V\r)^{k/m}}  {_2F_1} \l(\f{k}{2m}, \f{k}{2m}+1, \f{k}{m}+1, \f{4 U V}{(1+ U V)^2} \r),&U<0,\; V<0,\nonumber\\
& \phi= \frac{\eta}{f_{m,k}(r_c)}\;  \f{(-1)^k (2U)^{k/m} + (2 V)^{k/m}} {\l(1+ U V\r)^{k/m}}  {_2F_1} \l(\f{k}{2m}, \f{k}{2m}+1, \f{k}{m}+1, \f{4 U V}{(1+ U V)^2} \r) ,&U>0,\; V>0.
\eal

One can check that the above expressions are consistent with the general power series expansion described in \secref{sec:gen-grav}. Again, in generic situations, we find  power law divergences on the lightcones.

One can also analyze the behaviour of the Euclidean metric components $g(r,\theta),v(r,\theta)$ near $r=0$ based on the analysis in Appendix~\ref{sec:ads3_app}. The dependence on $r,\theta$ immediately gives us the dependence on $U,V$ in a near-horizon region of the Rindler wedges ($UV<0$) by analytic continuation. We find power-series expansions of the form
\bal
&g^{\pm}(U,V) = (-U V)^{k/m} \sum_{n=1}^\infty g^{\pm}_n (-U V)^{n}, \qquad g^0 (U,V) =  (-U V)^{k/m} \sum_{n=0}^\infty g^{0}_n (-U V)^{n} \\
&v^{\pm}(U,V) =(-U V)^{k/m} \sum_{n=0}^\infty v^{\pm}_n (-U V)^{n}, \qquad v^0(U,V) =   (-U V)^{k/m} \sum_{n=1}^\infty v^{0}_n (-U V)^{n}
\eal
where $g^{\pm,0}_n,v^{\pm,0}_n$ are coefficients found by solving the Euclidean equations of motion, \Eqref{gm} -- \Eqref{vp}, in a power series expansion near $r=0$. Given these solutions in the Rindler wedges, the by-now-familiar ansatz defined by \Eqref{eq:rindler} and \Eqref{eq:past}
 gives the solutions in the past and future wedges.  In particular, when $ UV >0$ we find the solutions have the following form
\bal\label{eq:gvseries}
&g^{\pm}(U,V) = (U V)^{k/m} \sum_{n=1}^\infty g^{\pm}_n (-U V)^{n}, \qquad g^0 (U,V) = (U V)^{k/m} \sum_{n=0}^\infty g^{0}_n (-U V)^{n} \\
&v^{\pm}(U,V) =(U V)^{k/m} \sum_{n=0}^\infty v^{\pm}_n (-U V)^{n}, \qquad v^0(U,V) = (U V)^{k/m} \sum_{n=1}^\infty v^{0}_n (-U V)^{n}
\eal
We can check this by plugging the stress tensor inside the horizon regions and solving for the metric components directly in Lorentzian signature.

\subsection{Divergence structure of fixed-area states}
\label{sub:divergence_structure}

As we saw in the examples above, for general sources at the Euclidean boundary, fixed-area states lead to power-law divergences on the codimension-2 conical defects. Similar divergences then also appear on the lightcones emanating from the HRT surface in the Lorentzian spacetime. We now generalize these results to arbitrary dimension and study the divergence structure of the general solution. 

\subsubsection{Euclidean signature}
As usual, we begin by studying the divergence structure in Euclidean signature. We consider Einstein-Hilbert gravity coupled to classical scalar fields.  We expect other classical matter fields to give similar results. 

In quasi-cylindrical coordinates
the metric near the codimension-2 defect can be always be written as \Eqref{near-defect-euc}, 
\bal
ds^2= dz d\bar{z} +& T \f{(\bar{z} dz - z d\bar{z})^2}{z \bar{z}} + h_{ij} dy^i dy^j + 2 i W_j dy^j (\bar{z} dz - z d\bar{z}),
\eal
where $T, h_{ij}, W_j$ are functions of all coordinates $(z,\bar{z}, y^i)$. The metric components have series expansions in terms of powers of $z^{1/m},\bar{z}^{1/m}$ and $z\bar{z}$ near $r=0$ as given in \Eqref{T-function}-\eqref{eq:matter}.

As noted before, in the presence of a $U(1)$ symmetry, all fields are smooth near $r=0$ and there is no singularity in any derivative of the fields. Furthermore, when $m=\frac{1}{n}$ for integer $n$ there is a smooth $n$-fold cover (as in the Lewkowycz-Maldacena discussion of  gravitational Renyi entropies \cite{Lewkowycz:2013nqa}), so again there are no power law divergences (even in the quotient). But more generally we will find divergences. Thus, in the following discussion, we consider a generic situation where $\frac{1}{m}$ is not an integer and where every coefficient in the power-series expansion of any metric component and matter field is non-zero.

Analyzing the Christoffel symbols we find
\bal\label{chris-sing}
\Gamma_{\mu \nu}^{\rho} \sim r^{1/m-1},
\eal
where we use the $\sim$ notation to keep track of the leading,  non-smooth term appearing in various quantities like metric components, derivatives of the matter field, etc. In the case $m>1$, the explicit $r^{1/m-1}$ term in fact represents the most singular term in the Christoffel symbols. For the Riemann tensor, there are terms involving the square of Christoffel symbols and terms involving the second derivatives of the metric. From \Eqref{chris-sing}, it is easy to see that the former terms at most are of order $r^{2/m-2}$. It turns out that the latter terms give the most singular terms in the Riemann tensor. In order to see this, let us  define
\bal\label{A-tensor}
A_{\mu\nu\rho \sigma} \equiv R_{\mu \nu \rho \sigma}  -\Gamma_{\lambda \mu\sigma} \Gamma^\lambda_{\nu \rho} + \Gamma_{\lambda \mu \rho} \Gamma^{\lambda}_{\nu \sigma}.
\eal
We find
\bal\label{Riemannzzb}
A_{z\bar{z} z \bar{z} } = T_{,z\bar{z}} + \frac{1}{2} \l(T \f{z}{\bar{z}} \r)_{,zz} + \frac{1}{2} \l(T \f{\bar{z}}{z} \r)_{,\bar{z}\bar{z}} \sim r^{\f{2}{m}-2},
\eal
where we have used
\bal
T = T_{110} (z \zb)^{1/m} + T_{101} \bar{z} z^{1/m+1} + T_{011} \bar{z}^{1/m+1} z+ \cdots.
\eal
Similarly we find
\bal
A_{z\bar{z} z j } = \f{1}{2} \l( i (W_j \bar{z})_{,z\bar{z}} + i (W_j z)_{,zz} - T_{,z i} - \l( T \f{\bar{z}}{z} \r)_{,\bar{z}i} \r) \sim r^{1/m-1},
\eal
and 
\bal
A_{z\bar{z} i j  }, A_{ijkl} \sim  r^{1/m},
\eal
while 
\bal
A_{aijk} \sim r^{1/m-1},
\eal
where $a=z,\bar{z}$. The most singular term comes from $h_{ij,ab}$ and is given by
\bal
A_{aibj} = \f{1}{2} \l( g_{aj,ib} + g_{ib,aj} - g_{ab,i j} - g_{ij,ab} \r) \sim r^{1/m-2}.
\eal
So in this case the most singular component of the Riemann tensor is
\bal\label{riemann-sing}
R_{aibj} \sim \f{r^{1/m}}{r^2}.
\eal
This behavior can be confirmed for the ten-dimensional Riemann tensor in the example of $AdS_3 \times S^3 \times T^4$ discussed in \secref{sub:ads_3}. In that case, the coefficient of \Eqref{riemann-sing} contains $ (\f{1}{m}-1)$ and therefore when  $m=1$ there is no singularity.

The degree of singularity in \Eqref{riemann-sing} naively implies that the Ricci tensor has similar singularities. For instance $R_{zz}$ contains terms like $h^{ij} \pa_z^2 h_{ij}$ which naively can be as singular as $r^{1/m-2}$. However, it was shown in Ref.~\cite{Dong:2019piw} for pure gravity with a cosmological constant that solving the Einstein equations sets 
\bal
h_{,000}^{ij} h_{ij,100} = h_{,000}^{ij} h_{ij,010} =0.
\eal
This means that due to the equation of motion, the leading term for the Ricci tensor must be of form $r^{2/m-2}$ which are less singular than terms in $R_{zizj}$. If the matter coupled to gravity is classical, we expect that $R_{\mu \nu} \sim T_{\mu \nu}$ and $T_{\mu \nu} \sim \pa_\mu \psi \pa_\nu \psi$ where $\psi \sim r^{1/m}$ near $r=0$. Therefore in this case, the most singular terms that scale as $r^{1/m-2}$ must be in the Weyl tensor and not the Ricci tensor. 

\subsubsection{Lorentzian signature}
Generalizing the analysis of divergences to Lorentzian signature is now straightforward. Continuing  $z \rightarrow V, \zb \rightarrow - U $, the Lorentzian metric takes the form
\bal
ds^2= -dU dV -& T \f{(V dU - U dV)^2}{U V} + h_{ij} dy^i dy^j + 2 i W_j dy^j (V dU - U dV),
\eal
where since $T$ vanishes on the horizons $V=0$ and $U=0$, we see that $U$ and $V$ asymptotically become affine parameters as one approaches either horizon. We now repeat the analysis of the components of the Riemann tensor. The main difference from the Euclidean analysis is that $U,V$ are independent. As a result, we consider the derivative of the metric as $U \to 0$ for a fixed $V$ (and similarly $V\to 0$ for fixed $U$). Note that  \Eqref{near-defect-euc} was originally an expansion for the metric in $r = \sqrt{z \zb}$. A point with a fixed $V$ and $U \to 0 $ eventually ends up in the small $UV$ region where  \Eqref{near-defect-euc} is a valid expansion. We use this expansion in the Lorentzian signature to find the leading singularity as $U \to 0$. However, the dependence on $V$ at fixed $V$ can be arbitrary and we only keep track of powers of $U$. 

We find:
\bal\label{lor_riemann_1}
&A_{VUVU} \sim U^{1/m - 1}, \qquad A_{VUVj} \sim U^{1/m},\qquad A_{VUUj} \sim U^{1/m-1}, \nonumber\\
&A_{UV i j} \sim U^{1/m} ,\qquad A_{ijkl} \sim U^{1/m}, \qquad A_{V ij k } \sim U^{1/m}, \qquad A_{Uijk} \sim U^{1/m-1}\nonumber\\
&A_{U i U j} \sim U^{1/m-2}, \qquad A_{U i V j} \sim U^{1/m-1},\qquad A_{V i V j } \sim U^{1/m}.
\eal
Therefore, the most singular components of the Riemann tensor are $ R_{UiUj}\sim U^{1/m-2}$.\footnote{As discussed previously, if $m= 1/n$ for integer $n$, the singular terms in the Riemann tensor vanish.}

Although the Riemann tensor itself can be divergent at the horizon (depending on $m$), the displacement of nearby geodesics passing through the horizon will be negligible. This can be seen if we integrate the geodesic deviation equation near $U=0$ twice to find the tidal displacement for nearby geodesics. The most singular term in the displacement goes as $U^{1/m}$ which vanishes for $U=0$. Thus, we see that fixed-area states have relatively mild divergences when working in the classical limit.

\section{Summary and Discussion} \label{sec:disc} 

Our work above studied the spacetime geometry intrinsic to fixed-area states at leading order in the bulk Newton constant $G$. While the saddle point geometries typically used to prepare such states contain conical singularities, they represent sources involved in the preparation and are not part of the fixed-area spacetime itself. Instead, the fixed-area spacetimes satisfy the usual equations of motion without conical singularities.

With either fine-tuning or enough symmetry, the fixed-area spacetimes can be completely smooth at leading order in $G$. More generally, however, derivatives of fields may diverge on null congruences fired orthogonally from the fixed-area surface. In particular, as described in \secref{sub:divergence_structure}, for states defined by cutting open Euclidean path integrals without a $U(1)$ symmetry, one typically finds the curvature tensor to diverge as $U^{1/m-2}$ as these null congruences are approached, where $U$ is the affine null parameter orthogonal to the null congruence and $2\pi m$ is the opening angle of the Euclidean saddle that prepares the fixed-area state. The singularities are integrable, meaning that the total tidal distortion experienced by freely-falling particles crossing the null congruence is finite. Thus such singularities need not necessarily destroy infalling observers and, in fact, so long as the coefficients of such singularities are small the effect on such observers can be negligible.

Importantly, in our example in \secref{sub:jt_gravity_matter} we found that the equations of motion could be solved in a manner that continues the solution beyond the power-law divergences on the lightcones of the HRT surface.   The resulting solution then had a large smooth region in both the past and future wedges. While these regions are harder to analyze in higher dimensional contexts, as in the three-dimensional analysis of section \ref{sub:ads_3} they will remain amenable to study via both standard perturbation theory and a near-horizon power series expansion.  This provides strong evidence that a large smooth region will continue to exist in both the past and future wedges.\footnote{Though a spacelike singularity may develop after some proper time since, even before fixing the area, we might consider a state that describes a black hole.} Furthermore, while the singularities we find at the horizons do in principle raise concerns regarding our control over the effect of any UV corrections on solutions in the past and future wedges, these concerns can be tamed by smearing out the HRT surface in the transverse directions and thus effectively introducing a UV cutoff.

Such singularities can be strong enough that they remove the spacetime from the realm in which the initial value problem for the Einstein-Hilbert gravity is well-posed. For example, in 3+1 dimensions standard such results require the curvature to be appropriately square-integrable \cite{Klainerman:2012wt,Klainerman:2012wy}. This is clearly violated for sufficiently large $m$. However, in our context this may not be a problem as we impose additional boundary conditions at the fixed-area surface (and, in effect, on the orthogonal null congruences) adapted from the Euclidean analysis of Ref.~\cite{Dong:2019piw}. These conditions are chosen to make the Einstein-Hilbert variational principle well-defined, and one may hope that they similarly repair the initial value problem. However, we leave the detailed study of such issues for future work.

Additional singularities will arise once we consider quantum corrections. One way to see this is to recall the example of a bifurcate Killing horizon in which the Euclidean saddle had a $U(1)$ symmetry. Such saddles were just familiar Euclidean black holes with conical singularities inserted at the horizon so that they could match boundary conditions with some period $\beta_0$ unrelated to the usual inverse temperature $\beta$ of the Lorentz-signature fixed-area geometry. At the quantum level, this clearly prepares quantum fields around this black hole in a state of inverse temperature $\beta_0$ which differs from the inverse Hawking temperature $\beta$. This is well-known to give a singular stress tensor at the black hole horizon, and in fact the special case $\beta_0 =\infty$ corresponds to the Boulware vacuum state for the black hole. The Boulware vacuum is much like the Rindler vacuum on Rindler space, for which the stress tensor features a quadratic divergence at the horizon \cite{Birrell:1982ix}. The associated back-reaction on the metric would then force the Ricci-tensor to be quadratically divergent as well, so that general integrated tidal distortions diverge logarithmically. This suggests that using fixed-area states beyond leading order in $G$ will require taming this divergence by smearing out the fixed-area surface along the orthogonal two spacetime dimensions; see also related comments in Ref.~\cite{Dong:2019piw}. This may also be related to issues regarding quantum corrections to HRT-areas seen in Ref.~\cite{Witten:2021unn}. We hope to return to further study of such quantum corrections in the future.

It would also be useful to generalize our results to include perturbative higher-derivative corrections and non-minimal couplings. In this context, the area is replaced by a more general geometric entropy functional \cite{Dong:2013qoa,Camps:2013zua}. Nevertheless, in the leading semiclassical approximation, states of fixed geometric-entropy states are again constructed by using Euclidean saddles with conical defects~\cite{Dong:2019piw}. The general arguments described here should thus go through in a similar fashion. In particular, there is a similar power-series expansion for metric quantities in a conical defect spacetime in higher-derivative theories \cite{Dong:2019piw}. This will again give power-series solutions in the past and future wedges just as described in \secref{sec:gen-grav}. However, an important difference in this case is that the power series expansion involves more singular terms. To resolve this, one may again consider a smeared version of the fixed geometric-entropy state in order to obtain reasonable initial data for Lorentzian evolution. It would be interesting to understand such solutions in greater detail in future.

A final open question involves the states where we fix the area of an HRT surface $\gamma_R$ that is anchored to an asymptotically AdS boundary (say, at the edges of a boundary subregion $R$). In this case, the area is divergent. While one can renormalize the HRT-area by subtracting its expectation value, fluctuations of this renormalized area remain divergent; see e.g. \cite{Marolf:2020vsi}. As a result, projecting onto a small window of HRT-area eigenvalues would remove the state from the CFT Hilbert space.\footnote{The situation is much like that for the operator $:\phi^2(x):$ in the theory of a free scalar field.} A useful notion of fixed-area state in this context will thus require the introduction of an appropriate boundary UV cutoff.

This then raises the question of how such UV issues will manifest themselves in the boundary-anchored versions of the calculations described in this work. One possibility is that, in the absence of a UV regulator, a singular shock will arise at the boundary anchors and will propagate into the bulk toward both past and future.  On the other hand, related UV concerns arise in the study of the 
flow by taking Poisson brackets with the HRT area operator \cite{Bousso:2020yxi,molly,xi}.  But in that context, at least in AdS$_3$, the behavior turns out to be milder.  Indeed, in that context    Ref.~\cite{molly} showed that the bulk itself remains smooth, and that the CFT singularity is dual only to a singularity in the manner that the bulk and boundary are attached. It thus seems likely that boundary-anchored fixed-area geometries will be similar. A better understanding of the area operator from the CFT perspective, perhaps along the lines of Ref.~\cite{Belin:2021htw}, may also be useful for understanding such issues.


\acknowledgments

This material is based upon work by XD, DM, PR, and ZW supported by the Air Force Office of Scientific Research under award number FA9550-19-1-0360. The work of DM, PR, and AT was also supported by a grant from the Simons foundation. Our work was also supported in part by funds from the University of California.

\appendix
\section{Scalar field solution via Fourier expansion}

\subsection{Massless Field}\label{sec:massless-fourier}

Let us now take a moment to construct the solutions in Eqs.~\eqref{sol-massless-r2}, \eqref{sol-massless-r4} directly, without recourse to analytic continuation. As is well-known, the space of solutions to the Klein-Gordon equation in 1+1 dimensional Minkowski space has a basis given by plane waves. A general solution $\phi(T,X)$ may thus be written in the form
\bal\label{massless-modes}
\phi(T,X) = \int_{-\infty}^{+\infty} a(\zeta) e^{ i (|\zeta| T - \zeta X)} d\zeta + \int_{-\infty}^{+\infty} a^\ast(\zeta)e^{- i (|\zeta| T- \zeta X)} d\zeta.
 \eal
 The initial condition is given by \Eqref{massless-ini-dat}:
 \bal
 \phi_0 (X) = 2X^{k/m} \Theta(X) + 2(-1)^k (-X)^{k/m} \Theta(-X), \qquad \l. \pa_T \phi \right|_{T=0} =0,
 \eal
which yields
\bal
 a^\ast(\zeta) = a(-\zeta),
 \eal
 and
 \bal
a(\zeta) =  \frac{\Gamma \left(\frac{k+m}{m}\right) \left| \zeta \right| ^{-\frac{k+m}{m}} \left(-\left((-1)^k+1\right) \sin \left(\frac{\pi  k}{2 m}\right)+(-i) \left((-1)^k-1\right) \text{sgn}(\zeta ) \cos \left(\frac{\pi  k}{2 m}\right)\right)}{2\pi}.
 \eal
 This result simplifies for even and odd $k$:
 \bal\label{fcoeff-even}
 &a(\zeta) = - \f{\Gamma(1+k/m) \sin(\f{\pi k}{2m})}{\pi} |\zeta|^{-k/m-1}, &k \in 2 \mathbb{Z}\\
 &\label{fcoeff-odd}a(\zeta) = i \f{\Gamma(1+k/m) \cos(\f{\pi k}{2m})}{\pi} {\rm sign}(\zeta) |\zeta|^{-k/m-1}, &k \in 2 \mathbb{Z}+1.
 \eal
Here in finding \Eqref{fcoeff-even} and \Eqref{fcoeff-odd}, we rotated the contour of integration to make the integrals convergent. For $k \in 2 \mathbb{Z}$, the field $\phi(U, V)$ is given by
 \bal\label{fmassless-sol-odd-1}
 \phi(U, V) &= \int_{-\infty}^{+\infty} \l( - \f{\Gamma(1+k/m) \sin(\f{\pi k}{2m})}{\pi} |\zeta|^{-k/m-1} \r) \l( e^{ i (|\zeta| T - \zeta X)} + e^{- i (|\zeta| T- \zeta X)} \r) d\zeta \nonumber\\
 &= |V|^{k/m} + |U|^{k/m},
 \eal
 where we used
 \bal\label{identity-1}
 \int_0^{+\infty} d\zeta \zeta^{-k/m-1} (e^{i \zeta V} + e^{-i \zeta V}) &= 2 |V|^{k/m} \cos(\f{\pi k}{2m}) \Gamma(-k/m).
 \eal
In \Eqref{identity-1}, an analytic continuation in $k/m$ is needed  to make sense of the integral.  Similarly for odd $k$ we find
 \bal\label{fmassless-sol-odd-2}
 \phi(T,X) &= \int_{-\infty}^{+\infty} \l( i \f{\Gamma(1+k/m) \cos(\f{\pi k}{2m})}{\pi} {\rm sign}(\zeta) |\zeta|^{-k/m-1} \r) \l( e^{ i (|\zeta| T - \zeta X)} - e^{- i (|\zeta| T- \zeta X)} \r) d\zeta \nonumber\\
 &= {\rm sign}(V) |V|^{k/m} - {\rm sign}(U) |U|^{k/m}.
 \eal
Thus, we see that both solutions agree with the solution we found previously by using appropriate analytic continuations.

\subsection{Massive Field}
\label{sec:massive_fourier}

In \secref{sec:massive}, we guessed the solution in the future wedge for the massive scalar field theory using the separation of variables and continuity of the solution. Here we derive the same solution directly by performing a Fourier transform.

Doing the Fourier transform directly on the initial data is not easy in this case. One way to go around this is to use the integral representation of the Bessel function,
\bal
I_\nu (x) = -i 2^{-\nu-1} x^\nu \int \f{\Gamma(s)}{\Gamma(s+1/2) \Gamma(1/2-s) \Gamma(1+\nu-s)} \l( \f{x^2}{4} \r)^{-s} ds,
\eal
where the integral runs over the imaginary line with a positive real part.

Using this representation, the calculation is almost the same as the massless case. For simplicity, let's set $k$ to be even, although the analysis is quite similar for odd $k$. Defining $V= T+ X, U= T-X, \nu \equiv  k/m$, we have
\bal
\phi(T=0, X)= 2 I_{k/m} (\mu |X|) = -i  \int \f{\Gamma(s)}{\Gamma(s+1/2) \Gamma(1/2-s) \Gamma(1+\nu -s)} 2^{2s-\nu}  |\mu X|^{\nu- 2s}  ds.
\eal
As a result,the Fourier coefficients are given by
\bal
&a(\zeta) = \f{1}{4\pi} \int_{-\infty}^{+\infty} \phi(T=0,X) e^{-i \zeta X} dX \nonumber\\
&=- \f{i}{2\pi}  \int \f{\Gamma(s)}{\Gamma(s+1/2) \Gamma(1/2-s) \Gamma(1+\nu -s)} 2^{2s-\nu} \mu^{\nu -2s}  \l(- \Gamma(1+\nu-2s) \sin(\f{\pi (\nu-2s)}{2}) |\zeta|^{-\nu+2s-1} \r) \nonumber\\
& = \int b(s) |\zeta|^{2s - \nu -1},
\eal
where $b(s) \equiv  i\f{\Gamma(1+\nu-2s) 2^{2s-\nu} \sin(\f{\pi (\nu-2s)}{2}) \Gamma(s)}{2\pi \Gamma(s+1/2) \Gamma(1/2-s) \Gamma(1+\nu -s)} \mu^{\nu -2s}  $. 
Therefore, the field is given by
\bal\label{fourier-massive}
\phi(T,X) = \int ds \, b(s) \int |\zeta|^{2s-\nu-1} \l( e^{ i \l( \sqrt{\zeta^2+ \mu^2 } T- \zeta X  \r)}  +e^{ -i \l( \sqrt{\zeta^2+ \mu^2 } T- \zeta X \r) } \r) d\zeta.
\eal
This integral is hard to do in general, so for a simpler case, we instead check that \Eqref{fourier-massive} reduces to \Eqref{sol-massive-r2} when $X=0$. In this case we have
\bal
\phi(T,0) &= \int b(s) ds \int |\zeta|^{2s - \nu -1} \l( e^{ i \l( \sqrt{\zeta^2+ \mu^2 } T  \r)}  +e^{ -i \l( \sqrt{\zeta^2+ \mu^2 } T \r) } \r) d\zeta \nonumber\\
& = \int b(s) ds \int \mu^{2s-\nu } |\sinh(y)|^{2s-\nu-1 } \l( e^{i \mu T \cosh(y)} + e^{- i \mu T \cosh(y)} \r) \cosh(y) dy \nonumber\\
&\begin{multlined}[b]
= 2\int b(s) ds \int \mu^{2s-\nu} 2^{s-\nu/2-1/2} \f{\Gamma(s-\nu/2)}{\sqrt{\pi}} \times\\
\qquad \qquad \times \l( - K_{s-\nu/2+1/2}(- i \mu T) (-i\mu T)^{-s+\nu/2+1/2}  - K_{s-\nu/2+1/2}( i \mu T) (i\mu T)^{-s+\nu/2+1/2} \r)
\end{multlined}
\nonumber\\
& = -2\int b(s) ds \int \mu^{2s-\nu}  \f{\Gamma(s-\nu/2)}{\sqrt{\pi}} \int d\tilde{s} \f{1}{4\pi i} \Gamma(\tilde{s})  \Gamma(\tilde{s} - s +\nu/2-1/2) \l[ \l(\f{i \mu T}{2}\r)^{ 1-2 \tilde{s}} + \l(\f{-i \mu T}{2}\r)^{ 1-2 \tilde{s}} \r].
\eal
The first simplification is that
\bal
\sqrt{\pi}\Gamma(1+\nu-2s) 2^ {2s-\nu} = \Gamma(\nu/2+1/2-s) \Gamma(\nu/2+1-s),
\eal
and 
\bal
\Gamma(\nu/2+1-s) \Gamma(s-\nu/2) = \pi/ \sin(\pi (s-\nu/2)). 
\eal
Using these identities and integrating over $s$ (and wrapping poles to the left), we find
\bal
&- \int b(s) \mu^{2s-\nu} \Gamma(s-\nu/2) \Gamma(\tilde{s} - s+\nu/2 -1/2) \nonumber=  -i  \f{\Gamma(1-\tilde{s}) \Gamma(\tilde{s} +\nu/2 -1/2) }{\Gamma(3/2 - \tilde{s} +\nu/2)}.
\eal
Defining $s_1= \tilde{s}-1/2$, and using $\Gamma(1/2-s_1) \Gamma(1/2+s_1) = \pi/\cos(\pi s_1)$ we have
\bal
\phi(T,0) &= -\int ds_1 \f{i}{2\pi} \f{\Gamma(1/2-s_1) \Gamma(1/2+s_1) \Gamma(s_1+\nu/2)}{\Gamma(1+\nu/2-s_1)} 2 \cos (\pi s_1) \l(\f{\mu T}{2}\r)^{-2s_1} \nonumber\\
& = -  \int ds_1 \f{i}{\pi} \f{ \Gamma(s_1+\nu/2)}{\Gamma(1+\nu/2-s_1)}   \l(\f{\mu T}{2}\r)^{-2s_1} = 2 J_{k/m}( \mu T),
\eal
as expected from by \Eqref{sol-massive-r2} by setting $V=T, U=T$ and considering even $k$. 

\section{Finding Solutions by Analytic Continuation in JT gravity} 
\label{sec:analytic}

\begin{figure}
    \centering
    \includegraphics[scale=0.5]{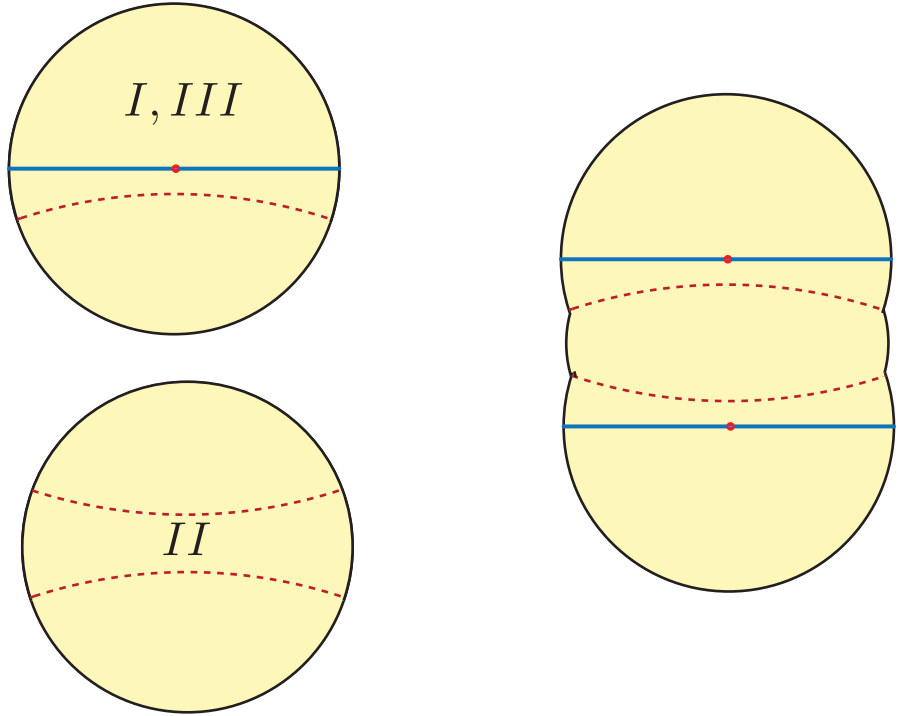}
    \caption{Left: We take solutions corresponding to a conical defect of opening angle $2\pi m$ and cut pieces of it, denoted I and III, slightly greater than half so as to avoid the defect. We then patch it onto a portion of the smooth $
    \text{AdS}_2$ solution, labelled II, by imposing matching conditions on the geodesic and at the asymptotic boundary. This leads to a regularized split solution which can be analytically continued to a Lorentzian spacetime. In the limit $\delta\rightarrow0$, we obtain the same solution that we expect from initial data evolution.}
    \label{fig:Reg}
\end{figure}

This appendix illustrates how the Lorentz signature solutions can be found by first regularizing the Euclidean conical singularities in a way that splits it (symmetrically) into two such singularities as in figure \ref{fig:split}, and then analytically continuing the region between them.

For a Euclidean JT gravity solution with a conical defect, we have 
\begin{equation}
ds^2=m^2(r^2-r_s^2)d\tau^2+\frac{dr^2}{r^2-r_s^2},
\end{equation}
\begin{equation}
\Phi=m \Phi_b r,
\end{equation}
where the periodicity of $\tau$ is $2\pi/r_s$. When $m=1$, the solution is smooth, so $r_s=\frac{2\pi}{\beta}$.

The boundary condition is that the boundary length is $ \frac{\beta}{\epsilon}$. This means that we should put the cutoff at $r=r_c=\frac{1}{m\epsilon}$, and the dilaton value there is $\frac{\Phi_b}{ \epsilon}$. 

We cut the solution at the following geodesic:
\begin{equation}
\begin{aligned}
r(\lambda) &=\frac{1}{m}\sqrt{\delta^{2}+(m r_{s})^{2}} \cosh \lambda, \\
\tau(\lambda) &=\frac{1}{ r_{s}} \arctan \left(\frac{m r_{s}}{\delta} \tanh \lambda\right),
\end{aligned}
\label{geod1}
\end{equation}
where $\delta $ indicates how far the geodesic is from the center of the disk. At $r=\frac{1}{m\epsilon}$, the affine parameter is 
\begin{equation}
\lambda_c= \cosh^{-1} \frac{1/\epsilon}{\sqrt{\delta^2+(m r_s)^2}}.
\end{equation}

In the middle region, we have a JT solution without a defect:
\begin{equation}
ds^2=(\bar r^2-\bar r_s^2)d\bar \tau^2+\frac{d\bar r^2}{\bar r^2-\bar r_s^2},
\end{equation}
\begin{equation}
\Phi=\bar \Phi_b \bar r,
\end{equation}
where the periodicity of $\bar \tau$ is $2\pi/\bar r_s$. We cut it at the geodesic
\begin{equation}
\begin{aligned}
\bar r(\bar \lambda) &=\sqrt{\bar \delta^{2}+\bar r_{s}^{2}} \cosh \bar \lambda, \\
\bar \tau(\bar \lambda) &=\frac{1}{\bar r_{s}} \arctan \left(\frac{\bar r_{s}}{\bar \delta} \tanh \bar \lambda\right).
\end{aligned}
\end{equation}
At the cutoff $\bar r=\bar r_c=\frac{1}{\epsilon}$, the affine parameter is 
\begin{equation}
\bar \lambda_c = \cosh^{-1} \frac{1/\epsilon}{\sqrt{\bar \delta^{2}+\bar r_{s}^{2}}}.
\end{equation}

We now glue these solutions together to regularize the conical defect solution to have a neighbourhood of a smooth solution near the $\mathbb{Z}_2$ symmetric slice. The first matching condition is that the affine parameters are the same:
\begin{equation}
\lambda_c=\bar \lambda_c,
\end{equation}
from which we get
\begin{equation}
\bar \delta = \sqrt{m^2 r_s^2-\bar r_s^2 +\delta^2}.
\end{equation}

The second matching condition is that the total boundary length is $\frac{\beta}{\epsilon}$:
\begin{equation}
4\left(\frac{\pi}{r_s}-\tau(\lambda_c)\right)+4\left(\frac{\pi}{2\bar r_s}-\bar \tau(\bar \lambda_c)\right)=\beta.
\end{equation}
To leading order in $\epsilon$, we know that the relation between $r_s$ and $\bar{r}_s$ can be solved from
\begin{equation}
\pi (r_s+\bar r_s)-2 \bar r_s \tan^{-1} \left(\frac{m r_s}{\delta}\right) -2 r_s \tan^{-1} \left(\frac{\bar r_s}{\bar \delta}\right) =0.
\end{equation}
This is in general hard to solve, but we can see that in the limit $\delta \to 0$, we have $\bar r_s=m r_s$. 

The last matching condition is that, the dilaton values should be the same at the boundary. Thus, we have 
\begin{equation}
\bar \Phi_b= \Phi_b.
\end{equation}

If we now analytically continue the middle region to Lorentzian signature, we obtain a smooth black hole with $\bar r_s=m r_s$, which means $\bar \beta =\frac{\beta}{m}$. The boundary dilaton is $\bar \Phi_b= \Phi_b$. Note that for this kind of spacetime that we constructed, the dilaton field only matches along the geodesic when we take $\delta \to 0$ limit. This is of course, as we expect from fixed-area state in pure JT gravity with $U(1)$ symmetry, identical to the microcanonical TFD at the temperature $\bar{\beta}$. 

More generally, this technique of regularizing the conical defect solution and analytically continuing from the neighbourhood of the $\mathbb{Z}_2$ symmetric slice might be useful. In certain situations, it may be simpler than solving the initial value problem.

\section{Solving the \texorpdfstring{AdS$_3$}{AdS3} metric perturbatively}\label{sec:ads3_app}
In \secref{sub:ads_3}, we solved for the scalar field in AdS$_3$ and described the solution for the backreacted metric. In this appendix, we write down all the equations for the metric components explicitly and show that they can be solved perturbatively as claimed in \secref{sub:ads_3}.

The solution for the scalar field to leading order is
\bal
\phi = 2\eta \cos( k \theta) \f{f_{m,k}(r)}{f_{m,k}(r_c)},\qquad f_{m,k}(r) = r^{k/m} {_2F_1}\l(\f{k}{2m}, \f{k}{2m}+1, \f{k}{m}+1, -r^2 \r).
\eal
From this solution, the stress tensor is decomposed into Fourier modes as
\bal
T_{\mu \nu} = \nabla_\mu \phi \nabla_\nu \phi - \frac{g_{\mu\nu}}{2} (\nabla \phi)^2 = T_{\mu \nu}^- e^{-2 i k \theta} + T_{\mu \nu}^+ e^{2 i k \theta} + T_{\mu\nu}^0.
\eal
Following \cite{Lewkowycz:2013nqa}, the ansatz for the metric to first order in $\eta^2$ is 
\bal
ds^2 = \frac{dr^2}{1+r^2 + g(r,\theta)} + m^2 r^2 d\theta^2 + m^2 (1+r^2)\l(1+ v(r,\theta)\r) dy^2,
\eal
where $g(r,\theta), v(r,\theta)$ are the metric perturbation, and have the following Fourier expansion,
\bal
g(r,\theta) = g^+(r) e^{2i k \theta} + g(r)^0+ g^-(r) e^{-2i k\theta},\qquad v(r,\theta) = v^+(r) e^{2i k \theta} + v^0(r)+ v^-(r) e^{-2i k\theta}
\eal
Plugging the stress tensor in the Einstein's equation (we set $8\pi G=1$)
\bal
R_{\mu \nu} - \frac{g_{\mu\nu}}{2} \l( R+2 \r) = T_{\mu \nu},
\eal
one finds that Fourier modes decouple to the first order. For the $yy$-components, we have
\bal
&4 k^2 g^{-}(r) + m^2 r(1+r^2)  g^{-\prime}(r) = 2 r^2 T^{-}_{yy} = \f{\eta^2}{f_{m,k}^2(r_c)} (1+r^2) \l(k^2 f_{m,k}(r)^2 - m^2 r^2 (1+r^2) f_{m,k}^{\prime\,2}\r),\label{gm}\\
&g^{0\prime}(r) = \frac{2 r}{m^2(1+r^2)} T^0_{yy} = \frac{\eta^2}{f_{m,k}^2(r_c)} \pa_r \l( r(1+r^2) f_{m,k}(r) f'_{m,k}(r) \r), \label{g0}\\
&4 k^2 g^{+}(r) + m^2 r(1+r^2)  g^{+\prime}(r) = 2 r^2 T^{+}_{yy} = \f{\eta^2}{f_{m,k}^2(r_c)} (1+r^2) \l(k^2 f_{m,k}(r)^2 - m^2 r^2 (1+r^2) f_{m,k}^{\prime \,2}\r)\label{gp},
\eal
where in \Eqref{g0}, the equation of motion for the scalar field is used. These equations are all first order differential equations and therefore, the solutions can be written in terms of integrals involving $f_{m,k}(r)$ and its derivatives. Using the solutions for $g^{\pm}(r),g^0(r)$, the $rr$ components of Einstein's equations give the differential equations for $v^{\pm},v^0$,
\bal
&-4k^2 (1+r^2) v^{-} + 2m^2 r^2 g^{-}(r) + m^2 r (1+r^2)^2 v^{-\prime}(r) = 2m^2 r^2 (1+r^2)^2 T_{rr}^{-}\label{vm},\\
& 2 r g^0(r) + (1+r^2)^2 v^{0\prime}(r) = 2 T^0_{rr} r(1+r^2)^2\label{v0},\\
& -4k^2 (1+r^2) v^{+} + 2m^2 r^2 g^{+}(r) + m^2 r (1+r^2)^2 v^{+\prime}(r) = 2m^2 r^2 (1+r^2)^2 T_{rr}^{+}\label{vp},
\eal
where 
\bal
&T^0_{rr} = \frac{\eta^2}{m^2 r^2 (1+r^2)f_{m,k}^2(r_c)} (-k^2 f_{m,k}(r)^2+ m^2 r^2 (1+r^2) f'_{m,k}(r)^2), \\
& T_{rr}^{-} = \frac{\eta^2}{2m^2 r^2 (1+r^2)f_{m,k}^2(r_c)} (k^2 f_{m,k}(r)^2 + m^2 r^2 (1+r^2) f'_{m,k}(r)^2),\\
& T_{rr}^{+} = \frac{\eta^2}{2m^2 r^2 (1+r^2)f_{m,k}^2(r_c)} (k^2 f_{m,k}(r)^2 + m^2 r^2 (1+r^2) f'_{m,k}(r)^2).
\eal
Therefore, the equations for $v^\pm(r), v^0(r)$ are also first order and the solutions can be written in terms of integrals of the stress tensor or power series expansions. We also checked that the power series solutions of $g^{\pm,0}, v^{\pm,0}$ satisfy other components of Einstein's equations and therefore, the verified the consistency of the metric ansatz. The form of the power series expansions have been listed in \Eqref{eq:gvseries}.
\addcontentsline{toc}{section}{References} 
\bibliographystyle{JHEP} 
\bibliography{references}

\end{document}